\documentclass[twocolumn,linenumbers,twocolappendix]{aastex631}

\usepackage{graphicx}

\newcommand{\myr}{\mbox{$M_{\odot}$~yr$^{-1}$}}
\newcommand{\myrkpc}{\mbox{$M_{\odot}$~yr$^{-1}$~kpc$^{-2}$}}
\newcommand{\Jyb}{\mbox{Jy beam$^{-1}$}}
\newcommand{\erg}{\mbox{erg s$^{-1}$ cm$^{-1}$}}
\newcommand{\SFRD}{\mbox{$\Sigma_{\mathrm{SFR}}$}}
\newcommand{\HII}{\mbox{\rm \ion{H}{2}}}

\newcommand{\Ha}{\mbox{H$\alpha$}}
\newcommand{\Hb}{\mbox{H$\beta$}}
\newcommand{\Paa}{\mbox{Pa$\alpha$}}
\newcommand{\Te}{\mbox{$T_{\rm e}$}}
\newcommand{\ffc}{\mbox{free-free continuum}}
\newcommand{\OIII}{\mbox{\rm [\ion{O}{3}]}}

\newcommand{\NII}{\mbox{\rm [\ion{N}{2}]}}

\newcommand{\SII}{\mbox{\rm [\ion{S}{2}]}}

\newcommand{\AV}{\mbox{A$_{V,R\mathrm{v}=3.1}$}}

\newcommand{\EBV}{\mbox{$E (B-V)$$_{R\mathrm{v}=3.1}$}}

\newcommand{\kVcal}{\mbox{$\mathrm{\mathrm{\kappa'}}$ ($V$)}}
\newcommand{\Rvcal}{\mbox{$R_V$}}
\newcommand{\Avcal}{\mbox{$A_V$}}

\newcommand{\Fukushima}{\affil{Physics and Mechatronics Course, Major in Symbiotic Systems Science and Technology, Graduate School of Symbiotic Systems Science and Technology, Fukushima University, 1 Kanayagawa, Fukushima, Fukushima, 960-1296, Japan}}
\newcommand{\NAOJ}{\affil{National Astronomical Observatory of Japan, 2-21-1 Osawa, Mitaka, Tokyo, 181-8588, Japan}}
\newcommand{\nihon}{\affil{Department of Physics, General Studies,
College of Engineering, Nihon University, Tamura-machi,
Koriyama, Fukushima 963-8642, Japan}}
\newcommand{\nagoya}{\affil{Institute for Space-Earth Environmental Research, Nagoya University, Furo-cho, Chikusa-ku, Nagoya, Aichi 464-8601, Japan}}
\newcommand{\alab}{\affil{Department of Physics, Graduate School of Science, Nagoya University, Furo-cho, Chikusa-ku, Nagoya, Aichi 464-8602, Japan}}
\newcommand{\Shizuoka}{\affil{Faculty of Global Interdisciplinary Science and Innovation, Shizuoka University, 836 Ohya, Suruga-ku, Shizuoka 422-8529, Japan}}
\newcommand{\joetsu}{\affil{Department of Geoscience, Joetsu University of Education, Yamayashiki-machi, Joetsu, Niigata 943-8512, Japan}}

\newcommand{\sokendai}{\affil{Astronomical Science Program, Graduate Institute for Advanced Studies, SOKENDAI, 2-21-1 Osawa, Mitaka, Tokyo 181-1855, Japan}}
\begin{document}

\title{Measuring 60-pc-scale Star Formation Rate of the Nearby Seyfert Galaxy NGC~1068 with ALMA, HST, VLT/MUSE, and VLA}

\correspondingauthor{Yuzuki Nagashima}
\email{s2380037@ipc.fukushima-u.ac.jp, yuzuki.nagashima@grad.nao.ac.jp}

\author{Yuzuki Nagashima}\Fukushima\NAOJ
\author[0000-0002-2501-9328]{Toshiki Saito}\NAOJ\Shizuoka
\author{Soh Ikarashi}\NAOJ\nihon
\author[0000-0001-6788-7230]{Shuro Takano}\nihon
\author[0000-0002-6939-0372]{Kouichiro Nakanishi}\NAOJ\sokendai
\author[0000-0002-6824-6627]{Nanase Harada}\NAOJ\sokendai
\author[0000-0002-8467-5691]{Taku Nakajima}\nagoya
\author[0000-0002-9695-6183]{Akio Taniguchi}\alab
\author[0000-0001-9016-2641]{Tomoka Tosaki}\joetsu
\author[0000-0001-9720-8817]{Kazuharu Bamba}\Fukushima

\begin{abstract}
Star formation rate (SFR) is a fundamental parameter for describing galaxies and inferring their evolutionary course. \HII\ regions yield the best measure of instantaneous SFR in galaxies, although the derived SFR can have large uncertainties depending on tracers and assumptions. We present an SFR calibration for the entire molecular gas disk of the nearby Seyfert galaxy NGC~1068, based on our new high-sensitivity ALMA 100~GHz continuum data at 55~pc ($=$~0\farcs8) resolution in combination with the {\it HST}~\Paa\ line data. In this calibration, we account for the spatial variations of dust extinction, electron temperature of \HII\ regions, AGN contamination, and diffuse ionized gas (DIG) based on publicly available multi-wavelength data. Especially, given the extended nature and the possible non-negligible contribution to the total SFR, a careful consideration of DIG is essential. With a cross-calibration between two corrected ionized gas tracers (\ffc\ and \Paa), the total SFR of the NGC~1068 disk is estimated to be 3.2~$\pm$ 0.5~\myr, one-third of the SFR without accounting for DIG (9.1~$\pm$ 1.4~\myr). We confirmed high SFR around the southern bar-end and the corotation radius, which is consistent with the previous SFR measurements.  In addition, our total SFR exceeds the total SFR based on 8~$\mu$m dust emission by a factor of 1.5.  We attribute this discrepancy to the differences in the young stars at different stages of evolution traced by each tracer and their respective timescales. This study provides an example to address the various uncertainties in conventional SFR measurements and their potential to lead to significant SFR miscalculations.
\end{abstract}

\keywords{Radio astronomy (1338) --- Seyfert galaxies (1447) --- Galaxy evolution (594) --- Star formation (1569) --- HII regions (694)}

\section{Introduction}\label{sec:intro}

Star formation rate (SFR) and its surface and volume density are fundamental observable parameters characterizing galaxies and their evolution \citep[e.g.,][]{KennicuttEvans12} and even the evolution of the Universe \citep[e.g.,][]{MadauDickinson14}. A spatial SFR distribution of galaxies provides key insights into where and how stars are formed in galaxies. As star formation plays an important role in the matter life cycle of galaxies by converting gas to stars and also spreading enriched materials to the surrounding interstellar medium (ISM) through stellar feedback processes, it is crucial to measure an accurate SFR distribution in galaxies in a spatially resolved manner.

Based on an assumed initial mass function (IMF) \citep[e.g.,][]{Kroupa01}, SFR can be estimated by measuring the amount of young massive stars through many different electromagnetic radiation mechanisms from X-ray to radio (see the review by \citealt{KennicuttEvans12}).
However, each of the current SFR measurement methods has many parameters with large uncertainties. Many previous works have derived  SFR values by making assumptions about these parameters \citep[e.g.,][]{Leitherer99,Calzetti08}. Examples of such parameters include electron temperature, extinction, and IMF when using SFR tracers emitted from \HII\ regions. SFR obtained from single wavelengths is subject to biases that depend on the characteristics of each wavelength \citep[e.g.,][]{Calzetti08, Calzetti13}. This was a major problem when comparing SFR obtained with other methods, as it creates a large difference and can cause significant misinterpretation of the SFR of each galaxy.

To alleviate uncertainties raised above, some studies have utilized multiple SFR tracers \citep[e.g.,][]{Calzetti08,Calzetti13} by calibrating emissions from massive stars and \HII\ regions, with cross-calibration among different ionized gas tracers \citep{Murphy11}. A key approach is using complementary tracers, e.g., hydrogen recombination lines and thermal bremsstrahlung emission \citep{Bendo15}. These tracer combinations effectively offset each other's weaknesses, enabling more accurate SFR measurements \citep[e.g.,][]{Bendo15, Bendo16, Michiyama20, MIlls21} with minimal bias and constraints of previously undetermined parameters like electron temperatures. Emissions from both tracers originate from \HII\ regions starting shortly after star formation, providing current star formation insights. The \Ha\ line, a powerful SFR tracer, faces extinction by interstellar dust, which is mitigated by using the Paschen line (\Paa), the strongest Hydrogen recombination line in near-infrared and less affected by dust \citep[e.g.,][]{Tateuchi13}. Conversely, thermal bremsstrahlung (free-free radiation) is observable by ALMA at millimeter wavelengths and is free from dust extinction, a unique advantage over other SFR tracers \citep{Draine11}. Despite being faint, recent ALMA observations have demonstrated its detectability in extragalactic systems with sufficient sensitivity \citep[e.g.,][]{Saito15,Kawana22}. 

When measuring the SFR of galaxies using an ionized gas tracer, it is necessary to exclude contributions from ionizing sources other than \HII\ regions. One such example is diffuse ionized gas (DIG). DIG has a lower ionization state than the typical \HII\ region \citep[e.g.,][]{Haffner09}. Although DIG had been recognized, detailed studies of the DIG were not actively pursued before state-of-the-art telescopes such as VLT/MUSE came online. Recent observational studies have thus begun to report DIG-corrected SFR \citep[e.g.,][]{Momose2013,Kaplan16,Kreckel16,Morokuma17,Belfiore22}. Furthermore, it has become common to derive SFR by taking into account DIG even in systematic surveys \citep[e.g.,][]{Kreckel16,Lacerda18,Tomicic21,Belfiore22}.

Another ionizing source, that could contaminate the measurement of an SFR, is an active galactic nucleus (AGN). Ionization by AGN produces more harder spectrum than young massive stars. This significantly affects the SFR measurement based on ionized gas tracers. To effectively separate AGN contributions, Baldwin-Phillips Terlevich (BPT) diagram \citep{Baldwin81} is a widely employed technique. This method mainly utilizes ratios of auroral lines (\OIII/\Hb\ and \NII/\Ha), which are highly sensitive to ionization sources, allowing for a more accurate determination of SFR.

\begin{figure*}[!th]
\begin{center}
\includegraphics[width=18cm]{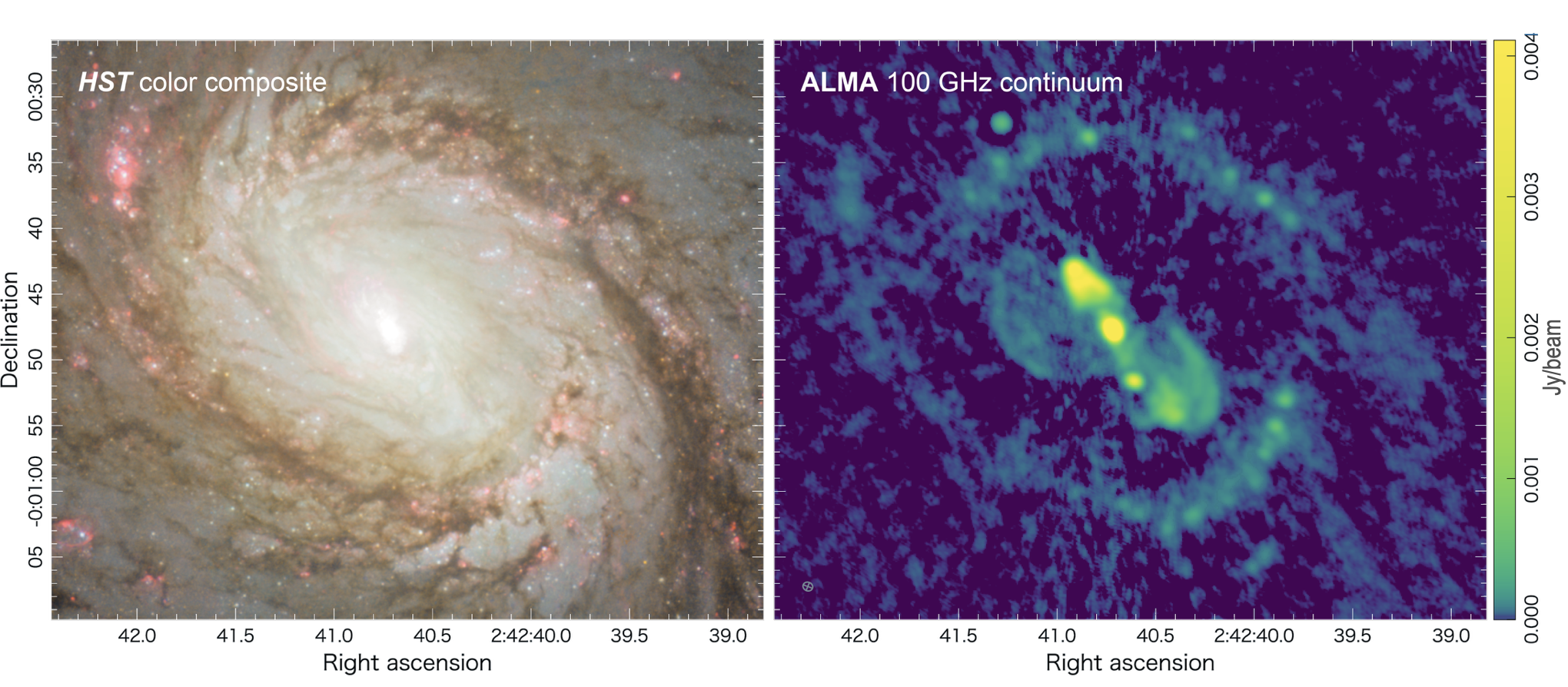} 
\end{center}
\caption{(left) {\it HST} color composite map of NGC~1068 (Credit: NASA, ESA \& A. van der Hoeven): WFPC2/$V$-band (cyan), ACS/$I$-band (yellow), and ACS/\Ha\ (red). (right) ALMA 100~GHz continuum map at 55~pc resolution.
}\label{Fig:opN1068_Fig1}
\end{figure*}

This study aims to obtain a high-quality map of the instantaneous SFR of the nearby prototypical spiral galaxy, NGC~1068 (Figure~\ref{Fig:opN1068_Fig1}; $D_{\rm L}$ $=$ 13.97~Mpc, z $\simeq$ 0.0038; 1\farcs0 $=$ 72~pc; \citealt{Anand2021}). NGC~1068 is known to host a biconical ionized and molecular gas outflow driven by the central AGN \citep[e.g.,][]{Das06,Garcia-Burrillo14,Garcia-Burillo2019,Saito22a}, but also known as one of the ideal targets to study star formation in external galaxies because it is nearly face-on \citep[$i$ $=$ 40$\degr$;][]{Planesas91}, massive \citep[$M_{\star}$ $\simeq$ 10$^{10.91}$ $M_\odot$;][]{Leroy19}, gas-rich ($M_{\rm H_2}$ $\simeq$ 1.5 $\times$ 10$^10$ $M_\odot$; \citealt{Planesas89}) and classified as a luminous infrared galaxy ($L_{\mathrm{IR}}$ $=$ 1$0^{11.4}$~$L_{\odot}$; \citealt{Armus09}). These properties make NGC~1068 easily observable and it became the subject of numerous observations and studies. Therefore, NGC 1068 serves as a laboratory-like galaxy with its rich observational data and its typical properties. By investigating the physical properties of this galaxy, we can enhance our understanding of galaxies at different ages or redshifts. 

The paper is divided into the following sections: 
In Section~\ref{sec:obs}, we describe the observations and the data used in this study. In Section~\ref{sec:method_result}, we present the method and the procedure to obtain a plausible SFR map using \Paa\ and \ffc. In Section~\ref{sec:disc}, we discuss the characteristics of star-forming activities and electron temperature variation in NGC~1068 in comparison with other galaxies in the literature. Finally, we summarize this paper in Section~\ref{sec:summary}.

\section{Observations and Data Processing}\label{sec:obs}

Here we briefly describe the data we use in this study, including ALMA, {\it HST}, MUSE, and VLA, and the hexagonal resampling as a common post-process after the imaging of each data.

\subsection{ALMA Data}\label{sec:obs:ALMA_Data}
\subsubsection{Band~3 Continuum}

We carried out two imaging spectral scans toward NGC~1068 with ALMA Band~3 and the 12-m array (2013.1.00279.S: PI=T. Nakajima; \citealt{Nakajima23}) and with the 7-m array (2021.2.00049.S: PI=T. Saito). To further gain sensitivity, we utilized ten projects publicly available in the ALMA archive that cover the Band~3 frequency coverage with the 12-m or 7-m array (2011.0.00061.S, 2012.1.00657.S, 2013.1.00055.S, 2013.1.00060.S, 2013.1.00221.S, 2015.1.00960.S, 2018.1.01506.S, 2018.1.01684.S, 2018.1.01684.S, and 2019.1.00130.S). This results in extremely wide-band observations covering nearly the entire Band~3 coverage from 84~GHz to 116~GHz (i.e., $\lambda$ = 3~mm). Total integration time with the 12-m array and the 7-m array are 24.9~hours and 26.3~hours, respectively.

Following \citet{Saito22a} and \citet{Saito22b}, we performed calibration with the observatory-provided calibration pipeline \citep{Hunter23} and imaging with the {\tt PHANGS-ALMA imaging pipeline} \citep{Leroy21a}. Spectral channels contaminated by strong molecular lines reported in \citet{Saito22b} and \citet{Nakajima23} are removed before imaging the continuum. We performed the continuum imaging with multi-term (multi-scale) multi-frequency synthesis (mtmfs; \citealt{Rau11}) using {\tt CASA tclean}. The reference frequency is set to be 99.7015~GHz, which is nearly the middle of the line-free frequency range. Hereafter, we call this Band~3 continuum map a 100~GHz continuum map (Figure~\ref{Fig:opN1068_Fig1} right). The details of the image properties are summarized in Table~\ref{table:ContPropTable}.

\subsubsection{Band~6 Continuum}
We collected eight Band~6 projects that employed the 12-m array from the archive (2013.1.00111.S, 2013.1.00188.S, 2013.1.00221.S, 2016.1.00023.S, 2016.1.00052.S, 2016.1.00232.S, 2017.1.01666.S, and 2018.1.00037.S). In addition to these archive data, we carried out a Band~6 spectral scan with the 7-m array (2021.2.00049.S: PI=T. Saito) that mostly covers between 211~GHz and 275~GHz to recover extended structures possibly missed by the archival 12-m data. We performed basically the same calibration and imaging procedures for the Band~3 data. We set the reference frequency to 260.5180~GHz when imaging with the mtmfs mode (hereafter 261~GHz continuum). The details of the image properties are summarized in Table~\ref{table:ContPropTable}.

\begin{figure*}[t]
\begin{center}
\includegraphics[width=17cm]{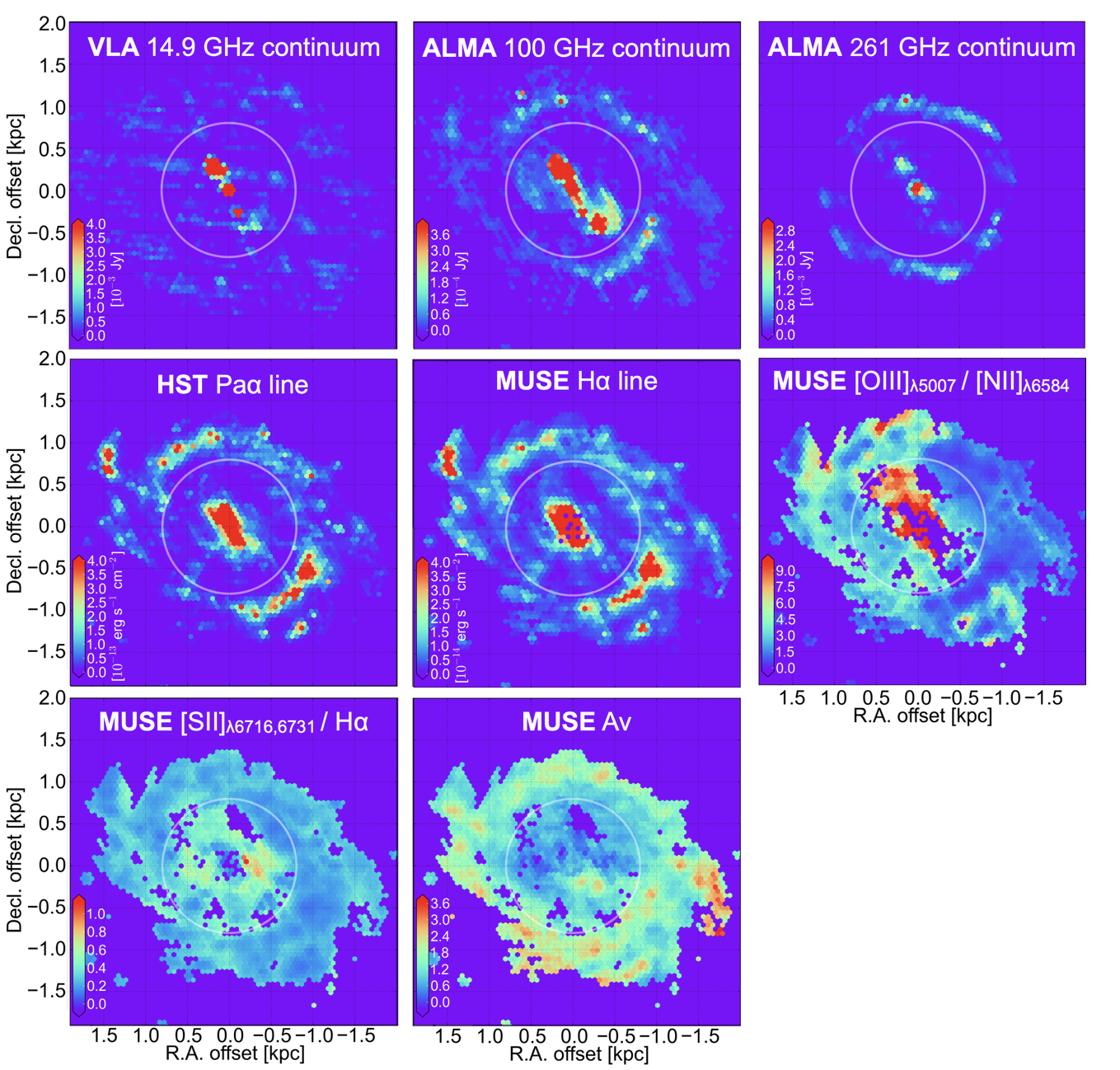} 
\end{center}
\caption{A gallery of multi-wavelength maps of NGC~1068 used in this study. All images are aligned into the same 55-pc-sized hexagonal grid after convolving the beam to 55~pc. Data points with $>3\sigma$ are displayed. The center position of these images is ($\alpha$, $\delta$)$_{\rm J2000}$ $=$ ($02^{\rm h}42^{\rm m}40\fs7091$, $-0^{\rm d}0^{\rm m}47\fs9450$). (top-left) VLA 14.9~GHz. (top-center) ALMA 100~GHz continuum. (top-right) ALMA 261~GHz continuum. The field of view (FoV) is nearly the central r$\sim$1.8~kpc. (middle-left) {\it HST} \Paa\ line \citep{Garcia-Burrillo14,Sanchez-Garcia22}. (middle-center) MUSE \Ha\ line \citep{Mingozzi19}. (middle-right) MUSE \OIII/\NII\ line ratio \citep{Mingozzi19}. (bottom-left) MUSE $\SII\lambda$6716,6731/\Ha\ line ratio \citep{Mingozzi19}. (bottom-center) MUSE extinction \citep[$A_{\rm V}$;][]{Mingozzi19}. The white circle highlights the central r$\sim$0.8~kpc region which is severely affected by the AGN jet and outflow \citep[e.g.,][]{Garcia-Burrillo14,Saito22b}.
}\label{Fig:maps_fig2}
\end{figure*}

\begin{deluxetable*}{ccccccccccc}
\tablecaption{Imaging properties of continuum data}
\label{table:ContPropTable}
\tablewidth{0pt}
\tablehead{
\colhead{$\nu_{\rm rest}$} & \colhead{$\theta_{\rm maj} \times \theta_{\rm min}$} & \colhead{RMS$_{\rm pix}$} & \colhead{RMS$_{\rm hex}$} & \colhead{Med $S_{\rm hex}$} & \colhead{S/N$_{\rm hex}$} & \colhead{$N_{\rm hex}$} \\
(GHz) & ($^{\prime}$$^{\prime}$) & (mJy beam$^{-1}$) & (mJy beam$^{-1}$) & (mJy) & (16$^{\rm th}$--50th$^{\rm th}$--84th$^{\rm th}$--max) & \\
(1) & (2) & (3) & (4) & (5) & (6) &(7) 
}
\startdata
14.9  & 0.46 $\times$ 0.38 & 0.2 & 2.1 & 1.3 & 0.47 --1.49--3.5--367.16 & 993 \\
99.7  & 0.76 $\times$ 0.66 & 0.01 & 0.1 & 0.02 & 3.6--6.6--20.9--1489.5    & 1236 \\
260.5 & 0.44 $\times$ 0.42 & 0.07 & 0.6 & 0.3 & 4.9--11.1--27.4--632.0  & 585
\enddata
\tablecomments{
Column 1: Continuum reference frequency.
Column 2: Major and minor axes of the original synthesized beam in full width at half maximum (FWHM) before convolution to 55~pc ($=$0\farcs8).
Column 3: Noise RMS per pixel.
Column 4: Noise RMS per hexagon.
Column 5: Median flux of detected hexagons.
Column 6: S/N ratio distribution of detected hexagons.
Column 7: Number of detected hexagons.}
\end{deluxetable*}

\begin{deluxetable*}{ccccccccccc}
\tablecaption{Imaging properties of line data}\label{table:lineProperties}
\tablewidth{0pt}
\tablehead{
\colhead{Line} & \colhead{$\lambda_{\rm rest}$} & \colhead{$\theta_{\rm res}$} & \colhead{RMS$_{\rm pix}$} & 
\colhead{RMS$_{\rm hex}$} & 
\colhead{Med $F_{\rm hex}$} & S/N$_{\rm hex}$ & $N_{\rm hex}$ \\
& (${\rm \AA}$m) & ($^{\prime}$$^{\prime}$) & (10$^{-15}$ erg cm$^{-3}$ s$^{-1}$) & (10$^{-15}$ erg cm$^{-3}$ s$^{-1}$) & (10$^{-15}$ erg cm$^{-3}$ s$^{-1}$) & (16$^{\rm th}$--50th$^{\rm th}$--84th$^{\rm th}$--max) & \\
(1) & (2) & (3) & (4) & (5) & (6) & (7) & (8)
}
\startdata
\OIII & 5007  & 0\farcs8 & 1.0 & 3.5 & 2.24 & 9.2--14.0--29.5--181.9 & 2180 \\
\Ha   & 6563  & 0\farcs8 & 0.16 & 0.6 & 5.39 & 21.4--53.8--100.6--306.6 & 2180 \\
\NII  & 6584  & 0\farcs8 & 0.5 & 1.9 & 3.63 & 15.0--27.2--44.1--155.6   & 2181 \\
\SII  & 6716  & 0\farcs8  & 0.02 & 0.08 & 0.1 & 14.6--27.6--48.5--243.0   & 2180 \\
\SII  & 6731  & 0\farcs8 & 0.02 & 0.08 & 0.1 & 14.0--24.1--38.8--183.1   & 2185 \\
\Paa  & 18756 & 0\farcs26 & 0.05 & 0.3 & 4.15 & 21.5--82.5--267.8--3018.9 & 1873 
\enddata
\tablecomments{
Column 1: Line name.
Column 2: Line rest wavelength.
Column 3: Typical seeing for MUSE and PSF size for HST ($^{\dagger}$\citealt{Mingozzi19}, \citealt{Sanchez-Garcia22}). 
Column 4:  Noise RMS per pixel.
Column 5: Noise RMS per hexagon.
Column 6: Median flux of detected hexagons.
Column 7: S/N ratio distribution of detected hexagons.
Column 8: Number of detected hexagons.}
\end{deluxetable*}

\begin{deluxetable}{lccccccccc}
\tablecaption{Photometry for the inner and outer regions of NGC~1068\label{table:PhotometryCenterOuter}}
\tablewidth{0pt}
\tablehead{
 & Unit & \colhead{$r < $ 0.8~kpc} & \colhead{$r > $ 0.8~kpc}
}
\startdata
$S_{\rm 14.9GHz}$          & Jy                                & 2.32$\pm$0.01   & $< $0.02 \\
$S_{\rm 100GHz}$           & Jy                                & 0.111$\pm$0.005 & 0.021$\pm$0.001 \\
$S_{\rm 261GHz}$           & Jy                                & 0.042$\pm$0.004 & 0.119$\pm$0.012 \\
\enddata
\tablecomments{
We only consider the statistical noise in this table.
}\end{deluxetable}

\subsection{Ancillary Data}\label{sec:obs:ancillary}

Here we briefly describe the ancillary data we use in this study.
\subsubsection{HST Data}\label{sec:obs:ancillary:HSTData}

We use the drizzled and processed {\it HST} \Paa\ map which is provided by M. S{\'a}nchez-Garc{\'\i}a (private communication). The details of the data properties and data reduction can be found in \citet{Garcia-Burrillo14} and \citet{Sanchez-Garcia22}. We conduct dust extinction correction by the Balmer decriment \citep{Calzetti00} to the \Paa\ map (See \ref{sec:method_result:extinctionCorr} for more details). The details of the image properties are summarized in Table~\ref{table:lineProperties}.

\subsubsection{MUSE Data}\label{sec:obs:ancillary:MUSE_Data}

We use the calibrated data cube of the integral-field spectroscopic data of NGC~1068 in the archive, collected with the MUSE \citep{Bacon10} on the VLT (program ID: 094.B-0321 (A), PI: Marconi). We downloaded the MUSE data cube from the ESO archive, as well as publicly available $\SII\lambda\lambda$9069,9532/$\SII\lambda\lambda$6717,6731 ratio map and dust extinction map \citep{Mingozzi19}.
The total integration time of the data used is 2800\,s. 
The data cube covers from 4750\,${\rm \AA}$ to 9350\,${\rm \AA}$ with a spectral resolution going from 1800 at 4650\,${\rm \AA}$ to 3750 at 9300\,${\rm \AA}$. 
The mean seeing for the observation is $\sim0''.8$.
Spectral extraction and measurement are performed using {\tt Astropy} \citep{Astropy13, Astropy18, Astropy22} and {\tt Specutils} \citep{Earl23}. 
The details of the image properties are summarized in Table~\ref{table:lineProperties}.

\subsubsection{VLA Data}\label{sec:obs:ancillary:VLA_Data}

We downloaded a 14.9~GHz (Ku-band) radio continuum map taken by VLA from the NRAO Data Archive. The retrieved data were originally taken with the BnA configuration with 27 antennas on December 5th, 1991, and reprocessed on July 29, 2008. The achieved spatial resolution is 0\farcs41 and the achieved sensitivity is 0.157~m\Jyb.

\subsection{Hexagonal sampling}\label{sec:obs:Hexagonal_sumpling}

We align all the maps used in this paper into the same hexagonal close‐packed grid \citep[e.g.,][]{Brok21,Saito22b} that naturally covers an area without gaps or overlaps \citep{Birch2007}. All images are resampled into 55 pc sized hexagonal grid pixels that are comparable to or lower than their spatial resolution, which ensures the independence of the spatial information of each pixel. Before the hexagonal regridding described below, we convolve all the ALMA and VLA maps to a round beam with 0\farcs8. That allows us to study star formation at each area covering a single or multiple \HII\ regions. We use the MUSE and {\it HST} maps as is whose spatial resolutions are comparable or better than 0\farcs8.  The size of each hexagon is set to be 0\farcs8 which minimizes the pixel-to-pixel correlation due to the beam oversampling when imaging. We perform this hexagonal regridding by averaging pixels over each hexagon. Then, clipping based on signal-to-noise ratio (at an S/N of 3) is done for further analyses. We evaluate the typical rms at signal-free hexagons of each map. Because the typical cloud size of the Milky Way is 40~pc in diameter \citep{Solomon87} which is comparable to the hexagon size (55~pc), we can regard each hexagon as representing at least one molecular cloud (see also \citealt{Sun18,Leroy21a}). Thus, in this paper, we aim to create a ``\textbf{giant molecular cloud (GMC)}-scale" SFR map, which will be used for further analysis in combination with molecular gas tracer data with \textbf{GMC}-scale spatial resolution (Nagashima et al. in preparation).

\section{SFR measurement}\label{sec:method_result}

In this section, we describe our method to derive a 55~pc-scale SFR map of NGC~1068.

\subsection{Method overview}\label{sec:method_result:Method_overview}

\begin{figure*}[t]
\begin{center}
\includegraphics[width=17cm]{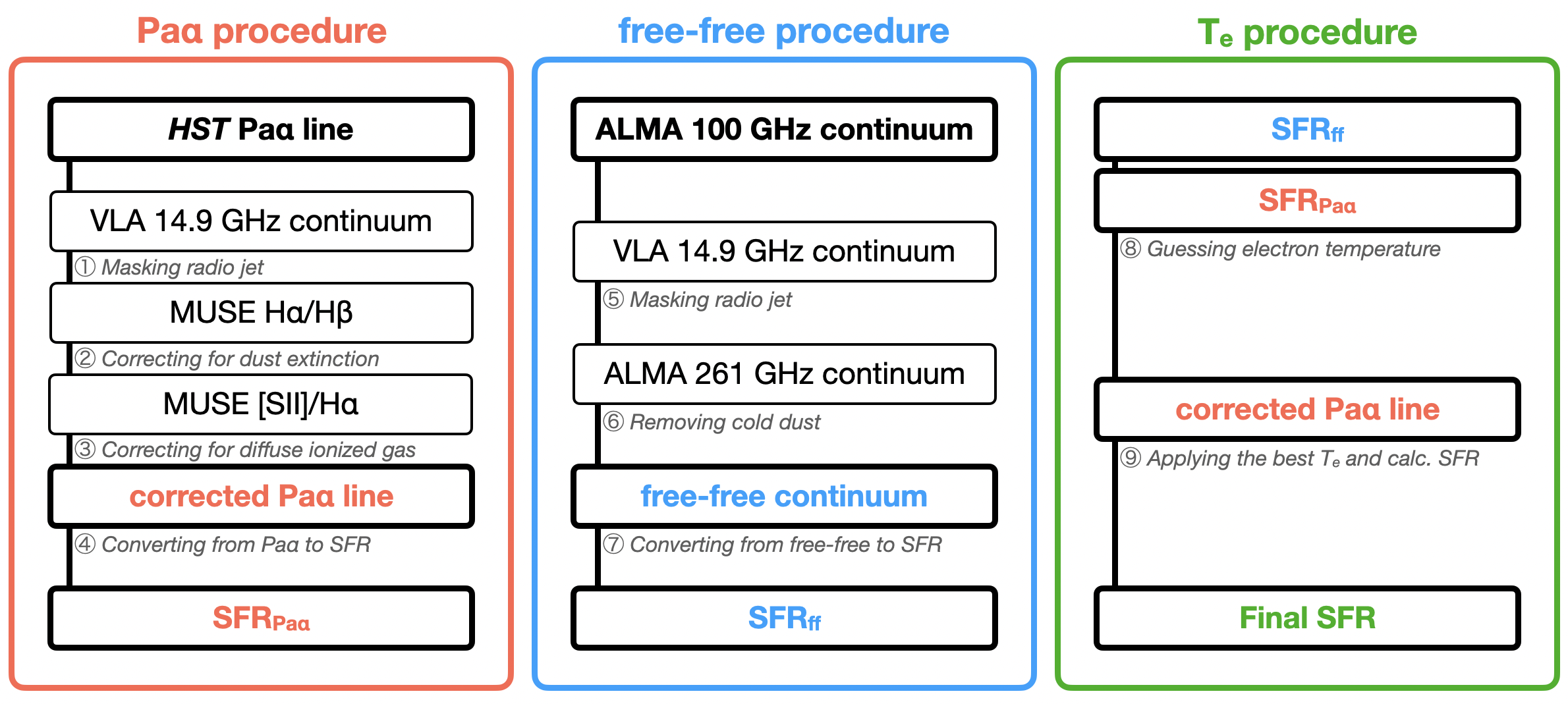} 
\end{center}
\caption{A graphical representation of the SFR calibration process. We start from two ionized gas tracer maps, {\it HST} Pa$\alpha$ \citep{Sanchez-Garcia22} and free-free continuum (this study) to obtain two independent SFR maps. Finally, we compare the two SFR maps to derive the best SFR map. Each step is described in Section~\ref{sec:method_result}.
}\label{fig:FlowChart_fig3}
\end{figure*}

As described in Section~\ref{sec:intro}, the interaction between free electrons and protons/ions in the ionized ISM is the emission mechanism of the thermal free-free continuum and hydrogen recombination lines. Therefore, SFR values derived from the two ionized gas tracers should agree with each other if all the other calibrations and corrections are properly applied \citep[e.g.,][]{Michiyama20}. Figure~\ref{fig:FlowChart_fig3} shows a flowchart describing the process of the SFR calibration employed in this study. Below, we describe each step.

\subsection{SFR prescription for Pa$\alpha$}\label{sec:method_result:SFRprescription}

The procedure to derive an SFR map from the calibrated {\it HST} \Paa\ map \citep{Garcia-Burrillo14,Sanchez-Garcia22} is listed as follows (see also the left panel of Figure~\ref{fig:FlowChart_fig3}):
\begin{itemize}
\item[1.] Masking AGN radio jet (Section~\ref{sec:method_result:MaskingAGN}).
\item[2.] Correcting for dust extinction (Section~\ref{sec:method_result:extinctionCorr}).
\item[3.] Correcting for diffuse ionized gas (Section~\ref{sec:method_result:DIGCorr}).
\item[4.] Converting from \Paa\ to SFR (Section~\ref{sec:method_result:convertToSFR}).
\end{itemize}
We describe this procedure step-by-step below. In Section~\ref{sec:method_result:Possible_AGN_contamination_to_ionized_gas}, we describe possible AGN contamination to the ionized gas tracers we employed.

\subsubsection{Masking AGN radio jet}\label{sec:method_result:MaskingAGN}

\begin{figure*}[t]
\begin{center}
\setlength{\fboxrule}{2pt} 
\includegraphics[width=16cm]{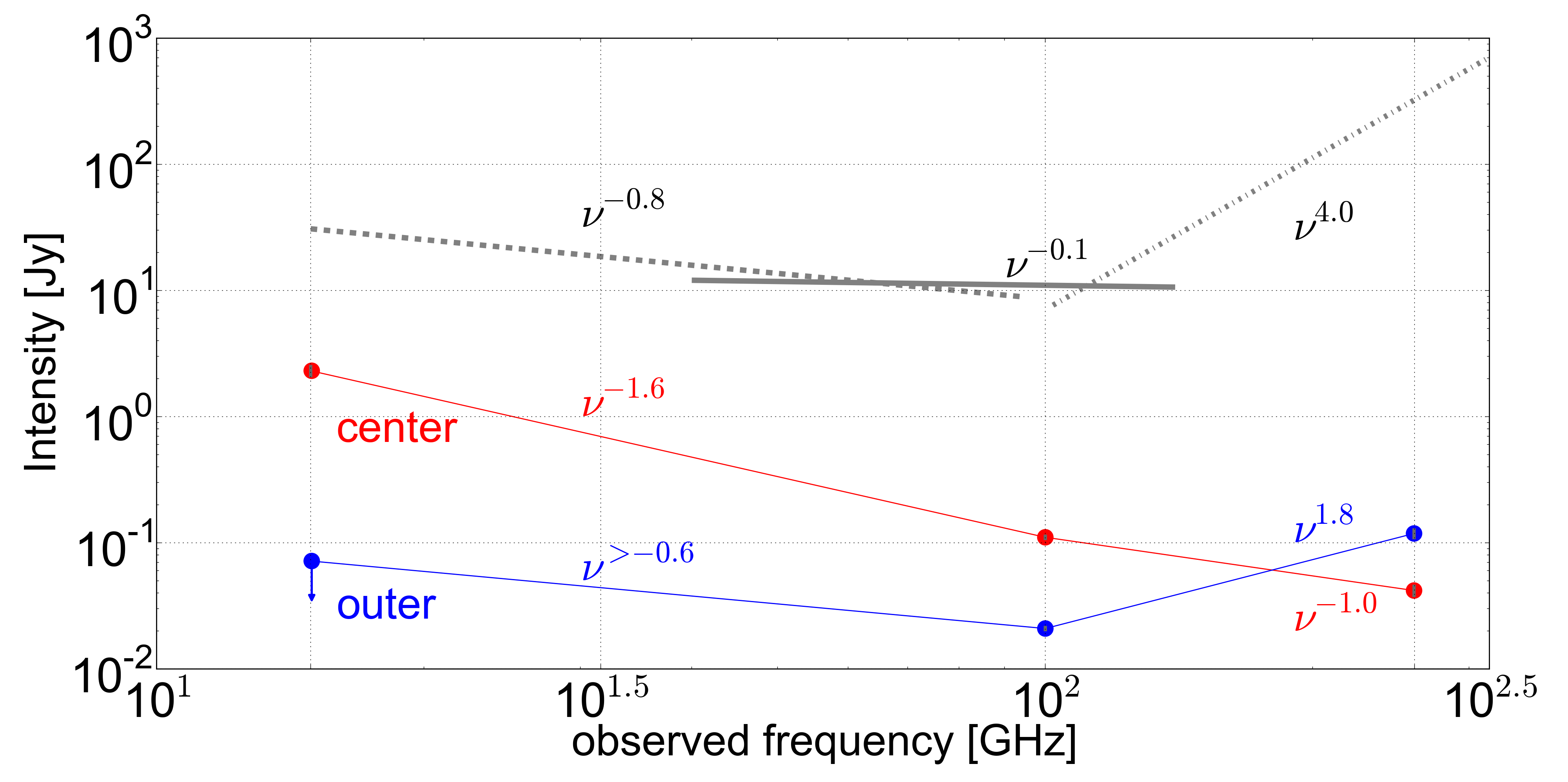} 
\caption{Observed radio-to-FIR spectral energy distributions of NGC~1068. SED of the central r$\sim$0.8~kpc is shown in red, and that of the outer part (i.e., starburst ring) is shown in blue. The black dotted line represents the typical slope of the synchrotron continuum ($=$ $-$0.8). The single-dotted line represents the typical slope of the Rayleigh-Jeans tail of the cold dust continuum ($=$ 4.0). The bold black line represents the typical slope of the free-free continuum ($=$ $-$0.1). The slope in the outer region is $<$ 4, implying that the 100 GHz flux is not dominated by thermal dust emission; therefore, the free-free flux can be estimated. Contrary, the inner region is likely dominated by synchrotron emission because of the observed relatively steep negative slope.}
\label{Fig:sed_fig4}
\end{center}
\end{figure*}

In general, synchrotron emission is the dominant component in continuum emission below 10 GHz. However, due to its relatively shallower spectral index of $-$0.8 \citep[e.g.,][]{Condon92,Gallimore04}, bright synchrotron emission can also significantly contribute to higher frequencies, even around 100 GHz.
One should remove the contribution of synchrotron emission from the 100~GHz continuum flux to accurately extract the \ffc. Especially, synchrotron emission from the AGN jet is known to be bright in the center of this galaxy \citep[e.g.,][]{Michiyama22}. This is demonstrated in Figure~\ref{Fig:sed_fig4}. Spectral energy distribution (SED) of the central r$\sim$0.8~kpc shows a decreasing trend, implying a dominant contribution from synchrotron emission. In the case of the disk, SED shows an opposite trend (e.g., free-free and thermal dust dominate). Therefore, we decided to mask the central r$\sim$0.8~kpc from further SFR analyses. This is because we can not accurately remove the contamination by synchrotron emission due to the lack of a sufficient number of SED data points.
Practically, we exclude hexagons detected with more than 5~sigma in the entire VLA 14.9~GHz continuum map that traces the bright radio jet components (Figure~\ref{Fig:maps_fig2}; see also \citealt{Gallimore04}).

In addition, we considered removing the possible contribution from the synchrotron emission in the starburst ring. However, the photometric measurements (Table~\ref{table:PhotometryCenterOuter}) imply that the effect of the synchrotron emission on the 100~GHz data is negligible in the starburst ring. For pixels at $r$ $>$ 0.8~kpc (i.e., starburst ring), we estimated the expected flux of the synchrotron continuum at 100~GHz to be $<$0.40 mJy when extrapolating the 3$\sigma$ upper limit from the flux of 14.9 GHz data to 100 GHz using a spectral index of  $\alpha$ = -0.8. This is $<$19$\%$ of the detectable flux (3$\sigma$ level) in the 100 GHz continuum in the starburst ring. Therefore, we conclude that synchrotron emission has negligible influence on the derivation of the free-free flux at the starburst ring.

\subsubsection{Correcting for dust extinction}\label{sec:method_result:extinctionCorr}

\begin{figure}[h]
\begin{center}
\includegraphics[width=8.5cm]{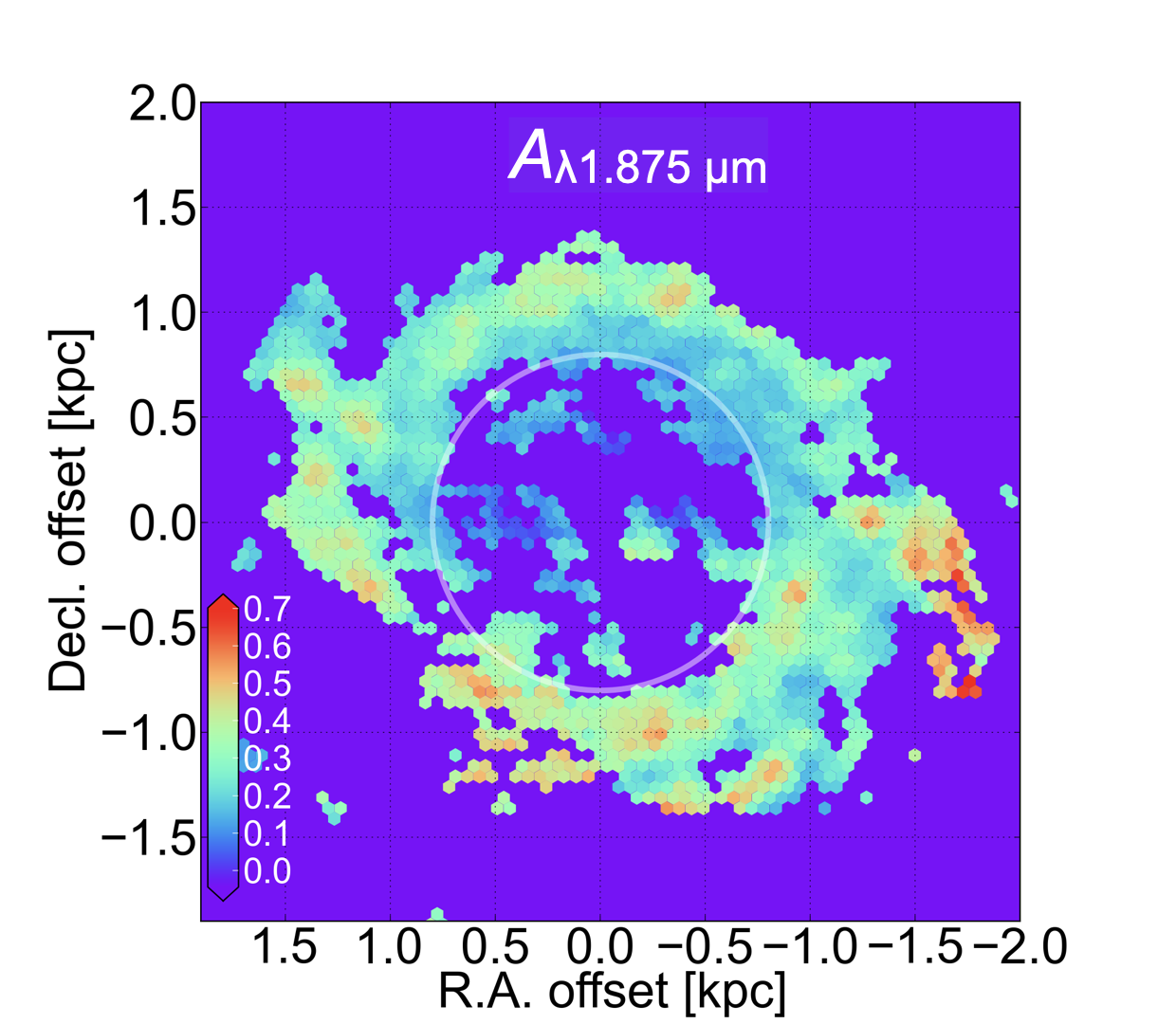}
\end{center}
\caption{Extinction map at 1.875~$\mu$m ($A_{\rm 1.875\mu m}$). The white circle highlights the central r$\sim$0.8~kpc region which is severely affected by the AGN jet and outflow \citep[e.g.,][]{Garcia-Burrillo14,Saito22b}. Data points with $>3\sigma$ are displayed.}
\label{Fig:A1.876mu_fig5}
\end{figure}

\begin{figure}[th]
\centering
\includegraphics[width=8.5cm]{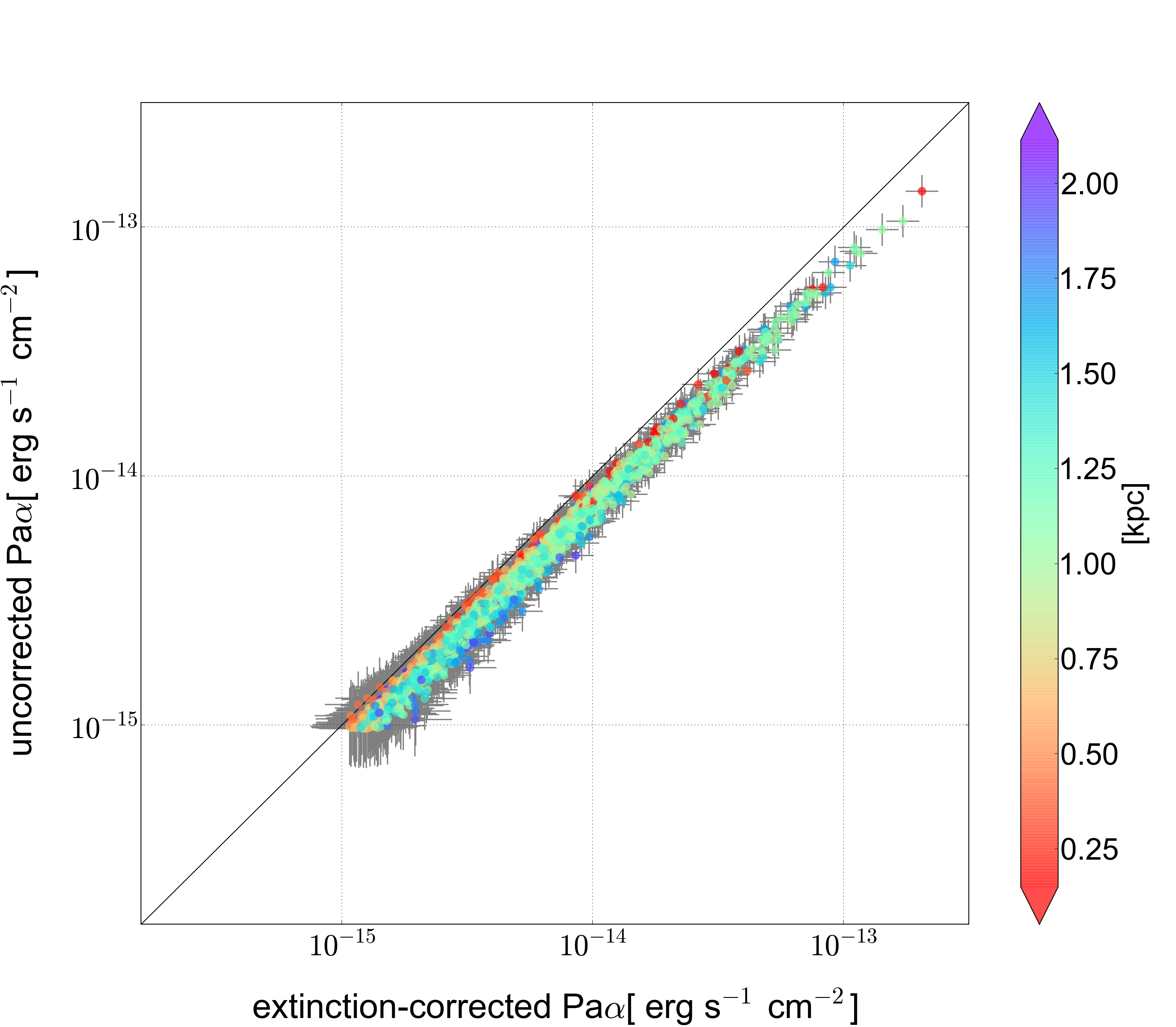}
\caption{A pixel-by-pixel comparison between extinction-corrected \Paa\ map and uncorrected \Paa\ map. The black line indicates a 1:1 relation. The color bar indicates the projected distance from the nucleus. Data points with $>3\sigma$ are displayed.}
\label{Fig:subExtinctionSc_fig6}
\end{figure}

We correct the \Paa\ map for dust extinction using an extinction map publicly available. This extinction map is one of the products from the MAGNUM survey \citep{Mingozzi19}, which was made by assuming a \citet{Calzetti00} extinction law with $R_{V}$ $=$ 3.1 (i.e., galactic diffuse ISM) and a fixed electron temperature $=$ 10$^5$~K. In our study, we recalculated the $A_{V}$ map by using $R_{V}$ $=$ 4.05, because \Paa\ at 1.875~$\mu$m does not meet the condition for Equation~7 of \citet{Calzetti00} (see Appendix~\ref{sec:appendix:Rv} for more details).  Figure~\ref{Fig:A1.876mu_fig5} is the map reflecting the extinction at 1.875~$\mu$m ($A_{\rm 1.875\mu m}$), the rest wavelength of \Paa. See Appendix~\ref{sec:appendix:Rv} for the conversion from $A_{V}$ to $A_{\rm 1.875\mu m}$. In Figure~\ref{Fig:subExtinctionSc_fig6}, we show a pixel-by-pixel comparison between dust-corrected and uncorrected \Paa\ maps. The extinction correction does not change individual pixel values significantly (less than $\simeq$0.2~dex), although it changes the total flux by $\simeq$25\%. This is comparable to or slightly higher than the uncertainty of the absolute flux calibration of the {\it HST} data \citep{Sanchez-Garcia22}. We note that our NGC~1068 data points follow the "Nearby galaxies" trend seen in the Av-dust relation compiled by \citet{Tomicic17}. We will discuss the dust properties of NGC 1068 in detail in a forthcoming paper (Nagashima et al. in prep).

\begin{figure}[h]
\centering
\setlength{\fboxrule}{2pt} 
\includegraphics[width=8.5cm]{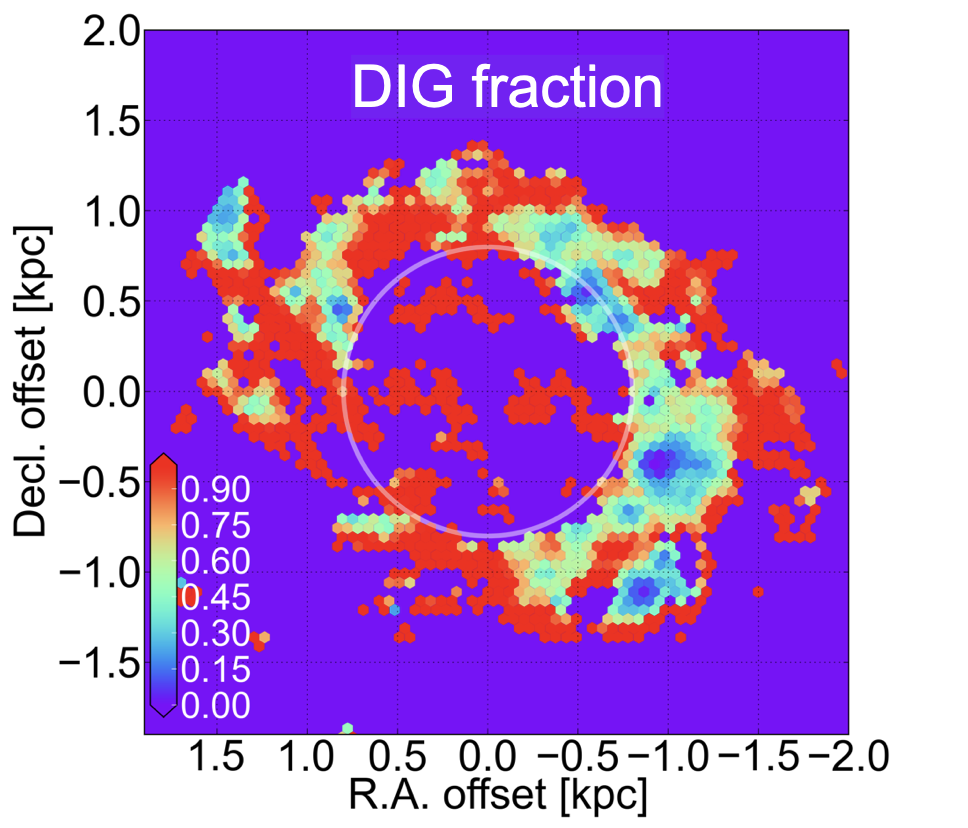}
\caption{DIG fraction map. Hexagons with 0 indicate non-detection of ionized gas tracer(s) or pure \HII\ region hexagons. The white circle highlights the central r$\sim$0.8~kpc region which is severely affected by the AGN jet and outflow \citep[e.g.,][]{Garcia-Burrillo14,Saito22b}. Data points with $>3\sigma$ are displayed. 
}
\label{Fig:rdig_fig7}
\end{figure}

\subsubsection{Correcting for diffuse ionized gas}\label{sec:method_result:DIGCorr}

In this study, we derive an SFR map with and without the DIG correction.

\begin{figure*}[th]
\centering
\includegraphics[width=0.9\textwidth]{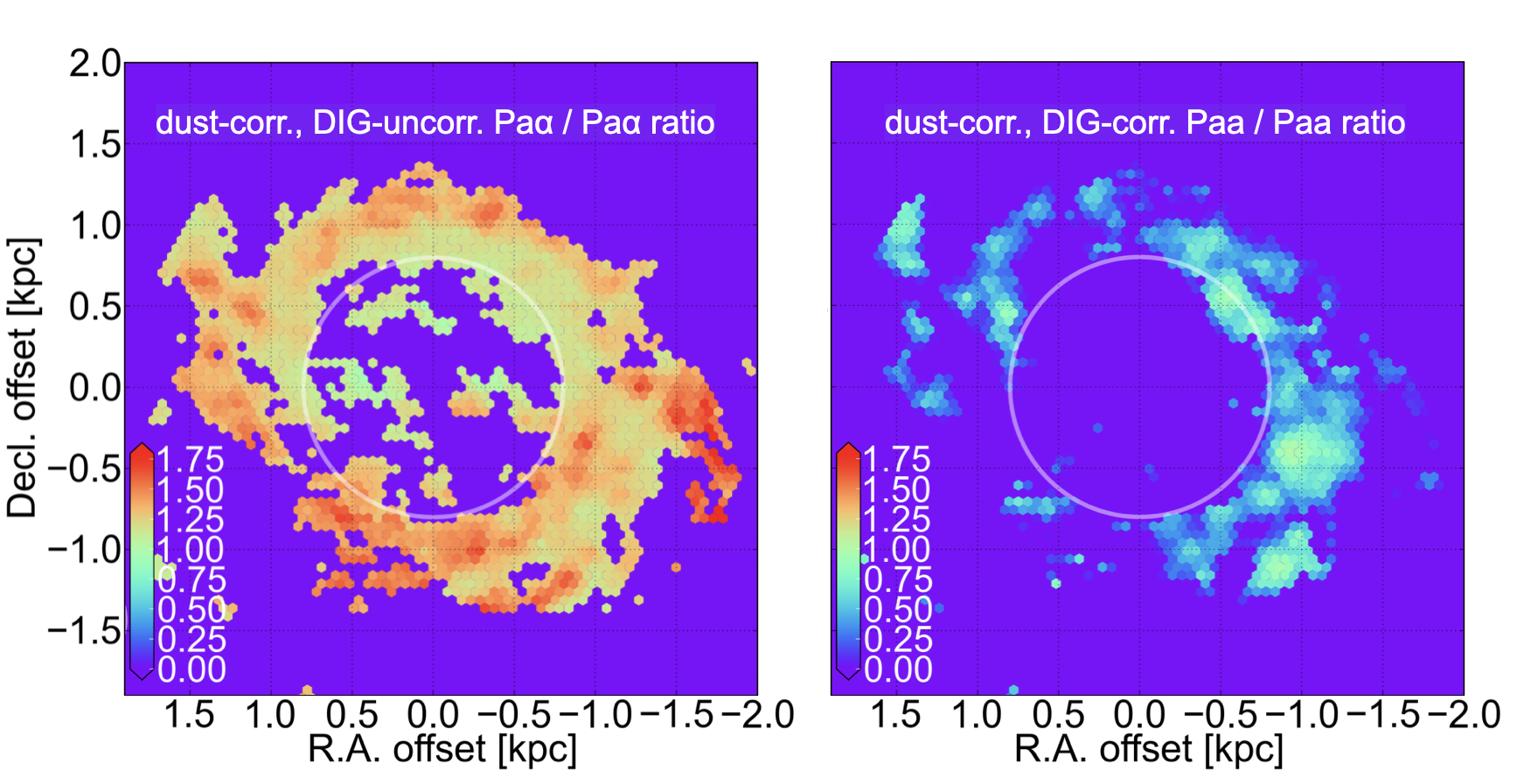}
\caption{Ratio maps of corrected to original data in NGC~1068 (left) A ratio map between the corrected dust extinction and the original \Paa\ map. (right) A ratio map between the corrected dust extinction and DIG and the original \Paa\ map. The white circle highlights the central r$\sim$0.8~kpc region which is severely affected by the AGN jet and outflow \citep[e.g.,][]{Garcia-Burrillo14,Saito22b}. Data points with $>3\sigma$ are displayed.}
\label{Fig:Paaflow_fig8}
\end{figure*}

We follow the DIG correction method described in \citet{Kaplan16}. The authors employed the optical \SII$\lambda$$\lambda$6716,6731/\Ha\ line ratio (hereafter \SII/\Ha; the lower-left panel of Figure~\ref{Fig:maps_fig2}) and \Ha\ flux to evaluate the DIG fraction at each pixel in \Ha\ maps \citep[see also][]{Kreckel16}. This method is expected to work because the typical DIG in the Milky Way is known to have a lower \Ha\ surface brightness and higher \SII/\Ha\ ratio compared to the typical \HII\ region \citep{Madsen06}. This method seems to also work for nearby galaxy data \citep[e.g.,][]{Kaplan16}, and thus we decided to employ this method in this study. After creating the DIG fraction map (Figure~\ref{Fig:rdig_fig7}), we apply this correction to the extinction-corrected \Paa\ map to subtract the DIG contribution. The \Paa\ maps corrected and uncorrected for DIG and extinction are shown in Figure~\ref{Fig:Paaflow_fig8}. The DIG-corrected map shows much less extended, low surface brightness structures implying that the method works as expected. Details of how we calculate the DIG fraction at each pixel of the \Paa\ map can be found in Appendix~\ref{sec:appendix:SIIHa}.

\subsubsection{Converting from \Paa\ to SFR} \label{sec:method_result:convertToSFR}

Here we describe the SFR prescription for the corrected \Paa\ line (see also \citealt{Murphy11,Bendo15,Bendo16,Michiyama20}). In this study, we follow the prescription described in \cite{Bendo16}. We use the Kroupa IMF \citep{Kroupa01}, which assumes a mass range of 0.1--100 $M_{\odot}$ and a metallicity (fraction of heavy elements to total elements) of 0.04, double the solar metallicity.

As SFR can be obtained by converting the number of ionized photons from massive stars, we start from the ionizing photon rate, $Q$,
\begin{equation}
\left (\frac{\mathrm{SFR}}{M_{\odot}\,\mathrm{yr}^{-1}}\right) = 5.41 \times 10^{-54}\left (\frac{Q}{\mathrm{s^{-1}}}\right).
\label{eq:1}
\end{equation}

$Q$ can be expresssed as,

\begin{eqnarray}
\lefteqn{\left (\frac{Q}{\mathrm{s^{-1}}}\right)} \nonumber \\
& = & \left (\frac{\alpha_{\mathrm{B}}}{\mathrm{cm^{-3}\ s^{-1}}}\right) \left (\frac{\epsilon}{\mathrm{erg\ s^{-1}\ cm^{-3}/}n_\mathrm{e}n_\mathrm{p}}\right)^{-1} \left (\frac{L_{\mathrm{RL}}}{\mathrm{erg\ s^{-1}}}\right),\nonumber\\ 
\label{eq:Q}
\end{eqnarray}

where $\alpha_{\rm B}$ is the total recombination coefficient, $\epsilon$ is the specific emissivity of each recombination line listed in \citet{StoreyandHummer95}, $n_{\rm e}$ is the electron volume density, $n_{\rm p}$ is the proton volume density (see \citealt{Michiyama20} for more details), and $L_{\rm{RL}}$ is the recombination line luminosity. With the case-B recombination assumption \citep{BakerandMenzel38}, $\alpha_{\rm B}$ varies with electron temperature (\Te) and electron density ($n_{\rm e}$). We use an interpolated relation between $\alpha_{\rm B}$ and $T_{\rm e}$ described in \citet{Michiyama20},
\begin{equation}
\left (\frac{\alpha_{\mathrm{B}}}{\mathrm{cm^{-3}\ s^{-1}}}\right) = 3.63 \times 10^{-10} \left (\frac{T_{\mathrm{e}}}{\mathrm{K}}\right)^{-0.79}.
\label{eq:a_B}
\end{equation}
With fixing $n_{\rm e}$ to be 10$^3$~cm$^{-3}$ and based on the list in \citet{StoreyandHummer95}, we employ the following relation between $\epsilon$ and \Te\ for \Paa,
\begin{equation}
\left (\frac{\epsilon}{\mathrm{erg\ s^{-1}\ cm^{-3}/}n_\mathrm{e}n_\mathrm{p}}\right) = 6.37 \times 10^{-24} \left (\frac{T_{\mathrm{e}}}{\mathrm{K}}\right)^{-1.05}.
\label{eq:eps}
\end{equation}
$\epsilon$ does not vary significantly within the range of $n_{\rm e}$ $=$ 10$^2$--10$^5$~cm$^{-3}$, while it varies within the range of \Te\ $=$ 1000--10000~K. We calculate $L_{\rm{RL}}$ by using the equation in \cite{Solomon05},
\begin{eqnarray}
\lefteqn{ \left (\frac{L_{\mathrm{RL}}}{\mathrm{erg\ s^{-1}}}\right) } \nonumber \\
& = & 4.0 \times 10^{30} \left (\frac{\int{f_{\mathrm{RL}}dv}}{\mathrm{Jy\ km\ s^{-1}}}\right) \left (\frac{\nu_{\mathrm{rest}}}{\mathrm{GHz}}\right) (1+z)^{-1} \left (\frac{D_{\mathrm{L}}}{\mathrm{Mpc}}\right)^2,\nonumber\\
\label{eq:L}
\end{eqnarray}
where $\int{f_{\mathrm{RL}}dv}$ is the line total flux, $\nu_{\mathrm{rest}}$ is the line rest frequency, z is the source redshift ($=$ 0.003793), and $D_{\rm L}$ is the source luminosity distance ($=$ 13.97~Mpc; \citealt{Anand2021}). Finally, \Paa-based SFR (SFR$_{\rm Pa\alpha}$) can be expressed as,
\begin{eqnarray}
\lefteqn{ \left (\frac{\mathrm{SFR}_{\mathrm{Pa}\alpha}}{M_{\odot}\ \mathrm{yr^{-1}}}\right) } \nonumber \\
& = & 5.41 \times 10^{-54} \left (\frac{\alpha_{\mathrm{B}}}{\mathrm{cm^{-3}\ s^{-1}}}\right) \left (\frac{\epsilon}{\mathrm{erg\ s^{-1}\ cm^{-3}}}\right)^{-1} \left (\frac{L_{\mathrm{RL}}}{\mathrm{erg\ s^{-1}}}\right) \nonumber \\ \label{eq:SFRPaa}\\
& \propto & \left (\frac{T_{\mathrm{e}}}{\mathrm{K}}\right)^{0.26} \left (\frac{L_{\mathrm{RL}}}{\mathrm{erg\ s^{-1}}}\right).
\label{eq:SFRPaa2}
\end{eqnarray}
This equation expresses that the \Paa-based SFR has a dependence of $T_{\rm e}^{0.26}$.

We note that the conversion coefficient can vary by a factor of two depending on assumed parameters and star formation history \citep[e.g.,][]{Bendo15}. The conversion coefficient in Equation~\ref{eq:1} were calculated using STARBURST99 \citep{Leitherer99}.

\subsubsection{Possible AGN contamination}\label{sec:method_result:Possible_AGN_contamination_to_ionized_gas}

Here we briefly describe the possible effect of the AGN outflow to the ionized gas tracers.

The AGN outflow in NGC 1068 is known to be spatially extended and bright in NGC~1068 \citep[e.g.,][]{Das06}. This outflow is a possible source of contaminating an SFR measurement using ionized gas tracers such as hydrogen recombination lines. 
To investigate the AGN effect on the SFR analysis we employed, we diagnosed it with a BPT diagram on a pixel-by-pixel basis \citep{Mingozzi19}. We classify whether the ionization source is starburst or AGN using the \SII/\Ha\ and \OIII/\Hb\ ratios. A small fraction of pixels in the northern part of the starburst ring (i.e., northern bar-end) is classified as pixels affected by the AGN outflow as shown in Figure~D.1 of \citealt{Mingozzi19}. However, this classification as AGN does not mean that the star formation rate derived for the molecular gas disk is contaminated. The orientation between the disk and the outflow reported in the literature suggests that the contamination is unlikely \citep[e.g.,][]{Das06,Garcia-Burrillo14,Saito22a}. Thus, the main body of the biconical ionized gas outflow is located in the extraplanar region and the ionized gas on the molecular gas disk is not directly and severely affected by the outflow except for the central kpc we masked out (as shown in Figure 17 of \citealt{Garcia-Burrillo14}). Therefore, we concluded that the northern part of the disk of NGC~1068 can be used for the SFR measurement.
Thus, we concluded that the northern part of the disk of NGC~1068 can be used for the SFR measurement. This is supported by the fact that the \Paa\ map shows quite similar structures seen in the 100~GHz continuum map, except for the central kpc region severely contaminated by the AGN outflow. Details of the analysis can be seen in \citealt{Mingozzi19}. 

\subsection{SFR prescription for 100~GHz continuum}\label{sec:method_result:100GHz_prescription}

The procedure to derive an SFR map from the ALMA 100~GHz continuum map is itemized as follows (see also the middle panel of Figure~\ref{fig:FlowChart_fig3}):
\begin{itemize}
\item[5.] Masking AGN radio jet (same as Section~\ref{sec:method_result:MaskingAGN}).
\item[6.] Removing cold dust (Section~\ref{sec:method_result:100GHz_prescription:Remove_colddust}).
\item[7.] Converting from \ffc\ to SFR (Section~\ref{sec:method_result:100GHz_prescription:Conv_from_100GHz}).
\end{itemize}
We describe this procedure step-by-step in this section.

\subsubsection{Removing cold dust continuum}\label{sec:method_result:100GHz_prescription:Remove_colddust}

\begin{figure*}[t]
 \begin{center}
\includegraphics[width=17cm]{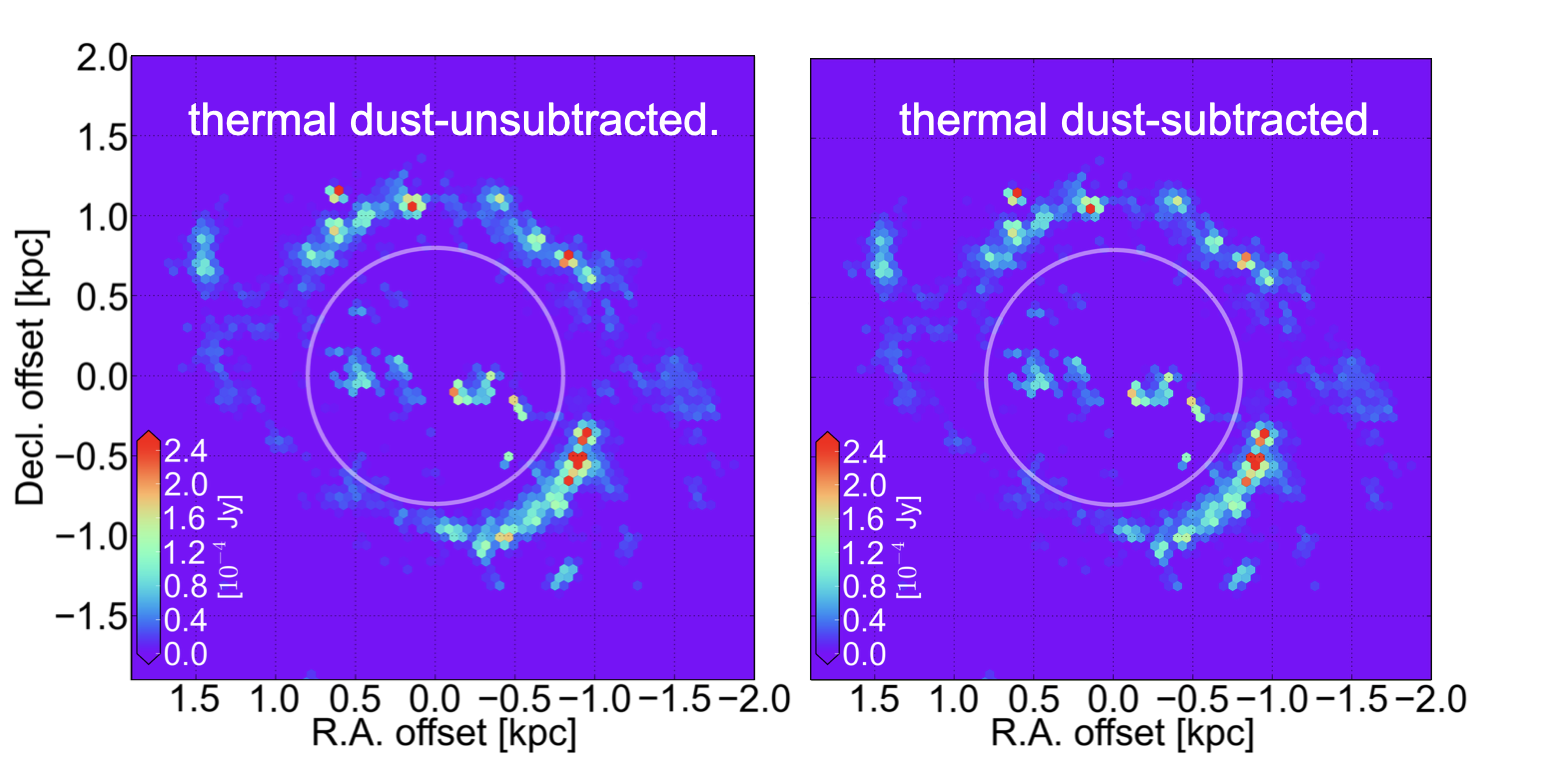} 
 \end{center}
\caption{100~GHz continuum maps of NGC~1068. (left) A map unsubtracted for the contamination from cold dust emission. (right) A map subtracted for the contamination. The white circle highlights the central r$\sim$0.8~kpc region which is severely affected by the AGN jet and outflow \citep[e.g.,][]{Garcia-Burrillo14,Saito22b}.
Data points with $>3\sigma$ are displayed.}\label{Fig:coldsub_fig9}
\end{figure*}

As described in Section~\ref{sec:method_result:MaskingAGN}, we decided to mask pixels detected within the central r$\sim$0.8~kpc of the VLA 14.9~GHz continuum map (Figure~\ref{Fig:maps_fig2} top-left), which results in removing the central jet structure from the 100~GHz continuum map. The remaining pixels are free from contamination by synchrotron emission. Thus, in this study, we consider correcting the 100~GHz continuum only for the cold dust contribution. we derive the expected cold dust flux map at 100~GHz by extrapolating the ALMA 261 GHz continuum map (Figure~\ref{Fig:maps_fig2} top-right) with a fixed spectral index of 4 and subtract it from the 100~GHz map. This index is a typical value for the Rayleigh-Jeans tail of nearby starburst galaxies \citep[e.g.,][]{Murphy11,Saito16}. The effect of this subtraction can be visually seen in Figure~\ref{Fig:coldsub_fig9}.

\begin{figure}[t]
\begin{center}
\includegraphics[width=8.5cm]{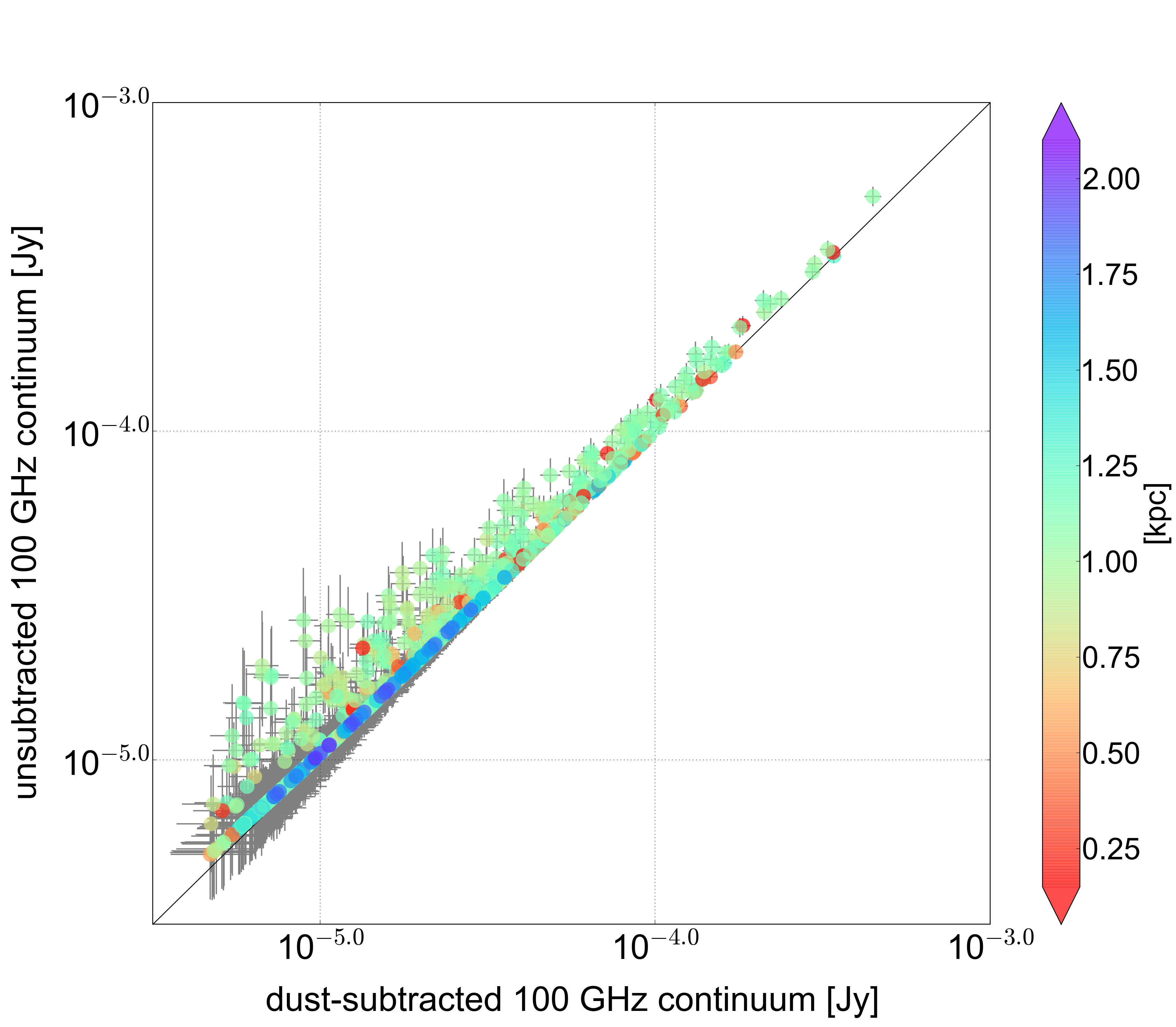} 
\end{center}
\caption{A pixel-by-pixel comparison between 100~GHz continuum map corrected for cold dust contamination and uncorrected one. The black line indicates a 1:1 relation. The color bar indicates the projected distance from the nucleus. Data points with $>3\sigma$ are displayed.
}\label{Fig:coldsub_sc_fig10}
\end{figure}

Figure~\ref{Fig:coldsub_sc_fig10} shows a pixel-by-pixel comparison between the two 100~GHz continuum maps. Most pixels distribute along the 1:1 relation (i.e., no flux change), although few pixels show $\lesssim$0.3~dex change. The difference in the total flux is 13$\pm$1~\%, highlighting the importance of the subtraction of cold dust component from the 100~GHz flux. It is not that large a percentage, but it is important to note that we cannot consider all of the Band 3 continuum flux coming from \ffc. We regard the dust-subtracted 100~GHz map as a \ffc\ map hereafter. Note that if we assume a shallower Rayleigh-Jeans slope of 3.8 instead, the total flux of the cold dust component at 100~GHz increases by $\simeq$4\%. This factor is smaller than the nominal flux accuracy of the ALMA Band~6 data (10\%), and thus we regard the difference due to the assumed slopes as negligible. This could be further investigated by adding more data points to the higher frequency regime. For pixels detected in the 100 GHz continuum map but not in the 261 GHz continuum map (See Figure~\ref{Fig:maps_fig2}), subtraction is not conducted.

\subsubsection{Converting from free-free to SFR}\label{sec:method_result:100GHz_prescription:Conv_from_100GHz}

Following \citet{Michiyama20}, free-free-based SFR (SFR$_{\rm FFC}$) can be expressed as,
\begin{eqnarray}
\lefteqn{\left ( \frac{\mathrm{SFR_{FFC}}}{M_\odot\ \mathrm{yr^{-1}}} \right)} \nonumber \\
& = & 9.49\times10^{10}g_{\mathrm{ff}}^{-1} \left (\frac{\alpha_{\mathrm{B}}}{\mathrm{cm^3\ s^{-1}}}\right) \left (\frac{T_{\mathrm{e}}}{\mathrm{K}}\right)^{0.5} \left (\frac{D_{\mathrm{L}}}{\mathrm{Mpc}}\right)^2 \left (\frac{f_{\mathrm{FFC}}}{\mathrm{Jy}}\right), \nonumber \\
\label{eq:SFRFFC} \end{eqnarray}
where $g_{\rm ff}$ is the gaunt factor and $f_{\rm FFC}$ is the free-free continuum flux. The gaunt factor can be written as,
\begin{equation}
g_{\mathrm{ff}} = 0.5535 \ln\left| \left (\frac{T_{\mathrm{e}}}{\mathrm{K}}\right)^{1.5} \left (\frac{\nu}{\mathrm{GHz}}\right)^{-1}Z^{-1} \right| - 1.682, \label{eq:gff}
\end{equation}
where Z is the ionic charge ($=$ 1). $g_{\rm ff}$ can vary by a factor of up to $\simeq$1.4 with 3000~K $<$ $T_{\rm e}$ $<$ 10000~K.

\subsubsection{Evaluating the effect of diffuse ionized gas on 100~GHz continuum}\label{sec:method_result:100GHz_prescription:DIG_100GHz}

Our SFR prescription for the \ffc\ does not include the DIG correction, unlike the prescription for \Paa. This is because our 100~GHz continuum map is not sensitive enough to detect \ffc\ from the DIG of NGC~1068. We estimate the expected \ffc\ flux from the DIG based on the observed \Paa\ flux of the DIG with some simple assumptions. The expected peak DIG flux at 100~GHz is 0.01~$\mu$Jy beam$^{-1}$ which is well below the sensitivity (0.9~mJy beam$^{-1}$). We refer the reader to Appendix~\ref{sec:appendix:contamiDIG} for a more detailed explanation and calculation.

Our conclusion can also be supported by the visual impression of the 100~GHz continuum map, i.e., the \ffc\ map shows much less extended structures compared to the original \Paa\ map or \Ha\ map that contains the DIG components (Figure~\ref{Fig:maps_fig2}).

\subsection{Derivation of the electron temperature}\label{sec:method_result:deriveTe}

Finally, the procedure to derive the best \Te\ value is itemized as follows (see also the right panel of Figure~\ref{fig:FlowChart_fig3}):
\begin{itemize}
\item[8.] Estimating \Te\ (Section~\ref{sec:method_result:TeDerive:TeGuess}).
\item[9.] Applying the best \Te\ and calculating SFR (Section~\ref{sec:method_result:TeDerive:SFRGuess}).
\end{itemize}

We describe this procedure step-by-step in this section.

\subsubsection{Estimating electron temperature}\label{sec:method_result:TeDerive:TeGuess}

\begin{figure*}[t]
\begin{center}
\includegraphics[width=18cm]{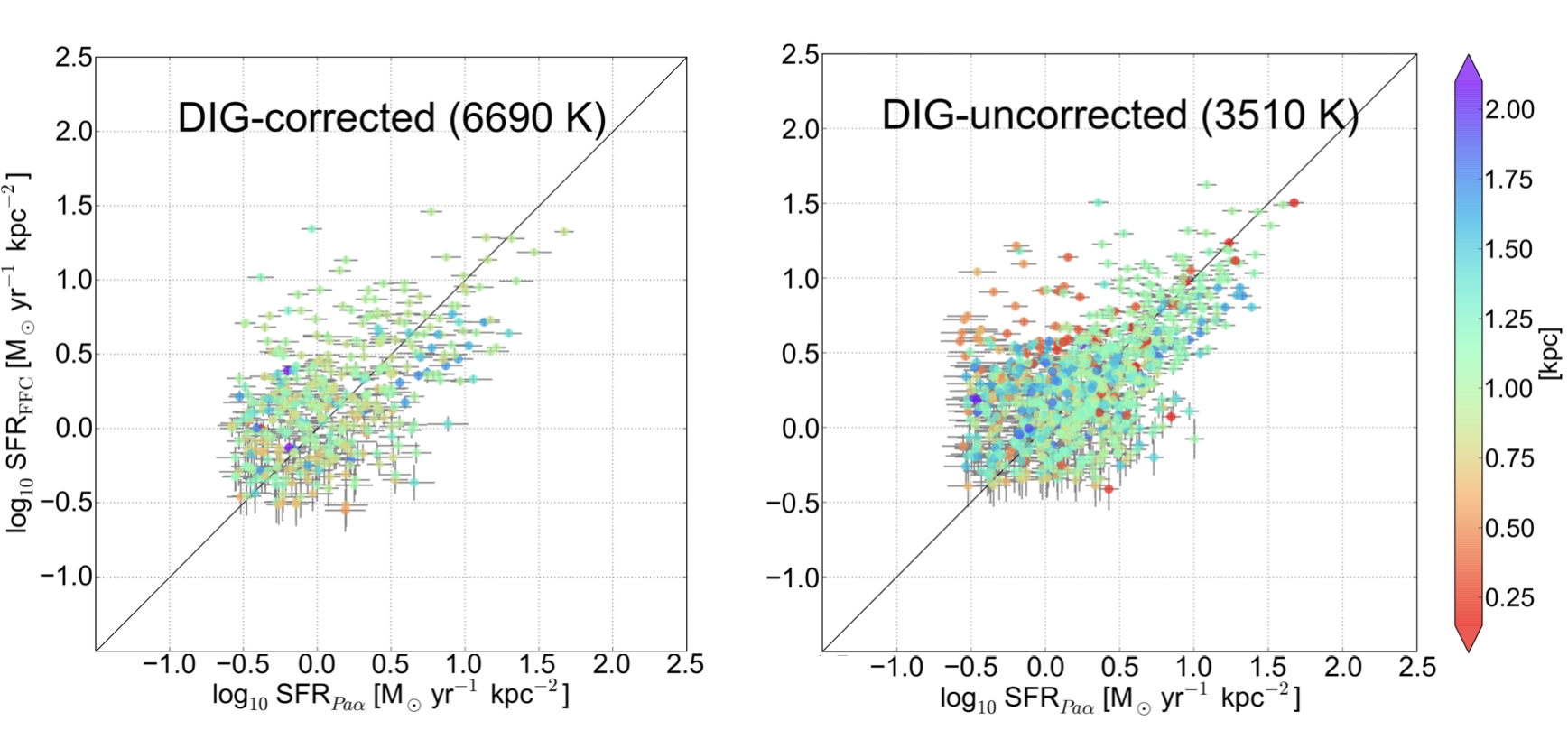} 
\end{center}
\caption{
Pixel-by-pixel comparisons between \Paa-based \SFRD\ map and free-free-based \SFRD\ map with best-fit \Te. The best-fit \Te\ are 6690 $^{+360}_{-340}$ and  3510 $^{+120}_{-110} $ for the case of DIG-corrected and DIG-uncorrected, respectively. The black line indicates a 1:1 relation. The color bar indicates the projected distance from the nucleus.
Data points with $>3\sigma$ are displayed.}\label{Fig:scatter_fig11}
\end{figure*}

\begin{figure*}[t]
\begin{center}
\includegraphics[width=18cm]{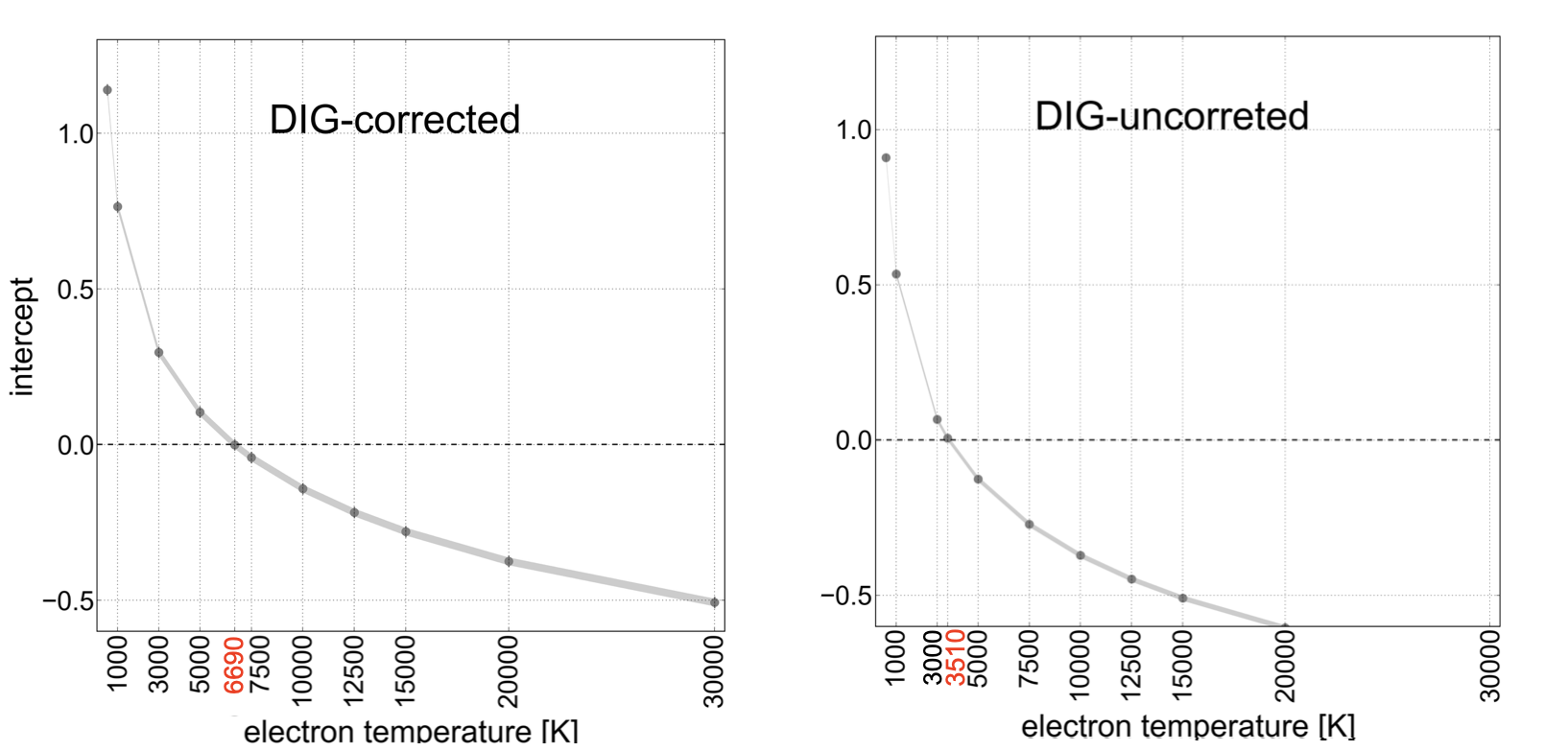} 
\end{center}
\caption{
Relations between the fitted intercept and \Te. (left) Plot for the DIG-corrected case. (right) Plot for the DIG-uncorrected case.
}\label{Fig:Te-in_fig12}
\end{figure*}

We derive two independent SFR maps based on two independent ionized gas tracers, \Paa\ and \ffc, after 55-pc-sized hexagonal gridding and corrections as explained in the previous Sections. Because the two tracers can be assumed to come from the same ionized gas ISM in galaxies, two SFR maps should agree with each other if all the other corrections and assumptions work (e.g., DIG correction and extinction correction). Any difference between the two SFR maps arises from different parameters used in Equations~\ref{eq:SFRPaa} and \ref{eq:SFRFFC}, i.e., \Te\ (see also \citealt{Michiyama20}). This is the concept of the \Te\ derivation employed in this study. In practice, we perform a least-squares regression with a fixed slope of unity for a scatter plot between the \Paa-based SFR map and the free-free-based SFR map (Figure~\ref{Fig:scatter_fig11}). When fitting, we weight data point located at ($x_{i}$, $y_{i}$) with 
\begin{equation}
w_{i} = \sigma_{i}^{-1} = \left( \frac{\sigma_{x,i}^{2}}{x_{i}^{2}} + \frac{\sigma_{y,i}^{2}}{y_{i}^{2}} \right)^{-0.5}
\end{equation}
where $\sigma_{i,{\rm x}}$ is the uncertainty of $x_i$. Any deviation of the intercept from 0 implies a wrong \Te\ value used for the SFR calculation. We search for the best \Te\ by varying the applied \Te\ value to calculate SFR and track the intercept (see examples in Figure~\ref{Fig:Te-in_fig12}). When all the corrections work and the assumed \Te\ value is correct, the intercept should become 0 (i.e., \Paa-based SFR and free-free-based SFR are matched). Note that this does not necessarily imply the total SFR derived from \Paa\ and \ffc\ agree because the number of detected pixels can be different.

We performed the regression for two cases: (1) free-free-based SFR and \Paa-based SFR corrected for DIG, and (2) free-free-based SFR and \Paa-based SFR not corrected for DIG. The searched \Te\ range is between 500 and 30000~K. The relation between \Te\ and the fitted intercept for cases 1 and 2 is shown in Figure~\ref{Fig:Te-in_fig12}. We regard the fitting error bar as the error of the intercept. Then, we calculate the error bar of \Te\ based on Figure~\ref{Fig:Te-in_fig12}.

In the case 1 (case 2), we derived \Te\ $=$ 6690$^{+360}_{-340}$~K (3510$^{+120}_{-110}$~K). Case 1 shows a higher value that can be naturally explained by the subtraction of \Paa\ flux due to the DIG correction.

\subsubsection{Applying the best \Te\ and calclating SFR}\label{sec:method_result:TeDerive:SFRGuess}

\begin{figure*}[t]
\begin{center}
\includegraphics[width=17cm]{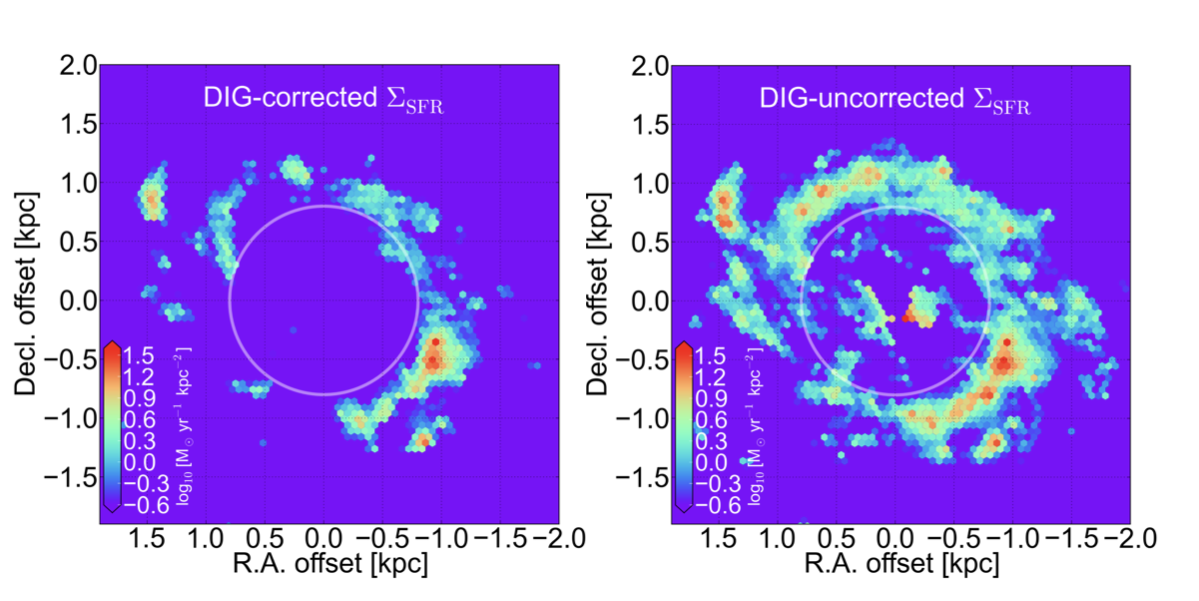} 
\end{center}
\caption{
\SFRD\ maps of NGC~1068. The best-fitted \Te\ derived in Section~\ref{sec:method_result:TeDerive:TeGuess} are applied to create these maps. (left) DIG-corrected \SFRD\ map (total SFR $=$ 3.2$\pm$0.5~\myr). (right) DIG-uncorrected \SFRD\ map (total SFR $=$ 9.1$\pm$1.4~\myr). The white circle highlights the central r$\sim$0.8~kpc region which is severely affected by the AGN jet and outflow \citep[e.g.,][]{Garcia-Burrillo14,Saito22b}. Data points with $>3\sigma$ are displayed.
}\label{Fig:SFRs_fig13}
\end{figure*}

Finally, we make \SFRD\ maps in the entire starburst ring of NGC~1068 (Figure~\ref{Fig:SFRs_fig13}) by applying the best-fitted \Te\ derived in Section~\ref{sec:method_result:TeDerive:TeGuess} to the \Paa\ map. As shown in Figure~\ref{Fig:SFRs_fig13}, The DIG correction significantly reduces extended, low surface brightness structures seen in the DIG-uncorrected \SFRD\ map. The total SFR is 3.2 $\pm$ 0.5~\myr\ for case 1 and 9.1 $\pm$ 1.4~\myr\ for case 2 in the entire starburst ring of NGC~1068. When correcting for DIG, \Te\ increases and the total SFR decreases. All measured parameters including \Te\ and SFRs are listed in Table~\ref{table:comparingSFRmap}. To make a fair comparison, we match the pixels between the case~1 and case~2 analysis, namely, we use pixels that survived in case~1 to calculate DIG-uncorrected \Te\ and SFR. We call this case 2b. We get \Te\ $=$ 2820~K, SFR $=$ 5.7 $\pm$ 1.2~\myr\ in this case. In case 2b, \Te\ is less than half of that in case~1, while SFR is about twice larger. 
Therefore, we concluded that the difference between case~1 and case~2 is not due to the difference in the number of pixels but is an actual trend due to the DIG correction.

\begin{figure*}[t]
\begin{center}
\includegraphics[width=18cm]{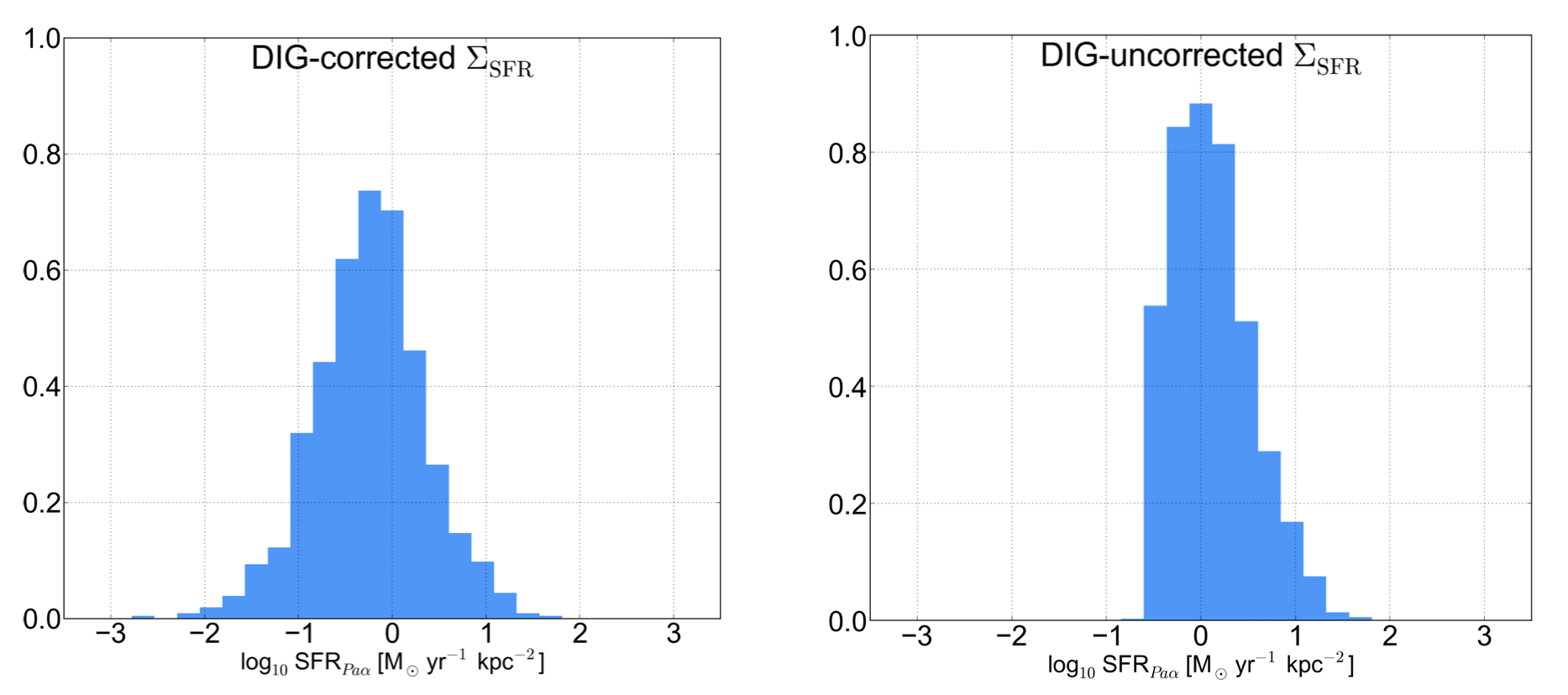} 
\end{center}
\caption{Normalized \SFRD\ histograms of the hexagons. (left) the histogram with DIG-corrected data. (right) the
histogram with DIG-uncorrected data. Data points with $>3\sigma$ are used here.
}\label{Fig:SFRh_fig14}
\end{figure*}

Histograms of the \SFRD\ are shown in Figure~\ref{Fig:SFRh_fig14}. The 16$^{\rm th}$--50$^{\rm th}$--84$^{\rm th}$--max percentiles are 0.2--0.6--2.1--46.7~\myrkpc\ and 0.5--1.2--3.5--47.1~\myrkpc\ for case 1 and case 2, respectively. As expected, the DIG-corrected \SFRD\ distribution shifts to the lower \SFRD\ regime compared to the DIG-uncorrected one as seen in the percentiles. In addition, different $T_{\rm e}$ values used for the SFR derivation contribute to these pixels. 

\begin{deluxetable*}{lccccccc}
\tablecaption{Calculated SFRs and their properties}\label{table:SFR}
\tablewidth{0pt}
\tablehead{
&\colhead{Best \Te} & \colhead{SFR$_{\rm total}$} & 
\colhead{SFR$_{\rm max}$} & 
\colhead{SFR$_{\rm median}$} & 
\colhead{SFR$_{\rm min}$} & $N_{\rm hex}$\\
& (K) & ($\rm{M_{\odot} yr^{-1}}$) & ($\rm{M_{\odot} yr^{-1} kpc^{-2}}$) & ($\rm{M_{\odot} yr^{-1} kpc^{-2}}$) & ($\rm{M_{\odot} yr^{-1} kpc^{-2}}$) &  \\
 & (1) & (2) & (3) & (4) & (5) & (6) }

\startdata
DIG-corrected case & 6690$^{+360}_{-340}$ & 3.2$\pm$0.5 & 46.7$\pm$7.0 & 0.6$\pm$0.1 & 0.002$\pm$0.0005 & 864 \\ 
DIG-uncorrected case & 3510$^{+120}_{-110}$ & 9.1$\pm$1.4 & 47.1$\pm$7.1 & 1.2$\pm$0.2 & 0.25$\pm$0.09& 1548
\enddata
\tablecomments{
Column 2: Total SFR of detected hexagons.
Column 3: Maximum SFR.
Column 4: Median SFR.
Column 5: Minimum SFR.
Column 6: Number of detected hexagons.
} \label{table:comparingSFRmap}
\end{deluxetable*}

\section{Discussion} \label{sec:disc}

\subsection{Reliability of this SFR prescription in comparison with other studies}\label{sec:disc:sfr_tsai}

A previous study by \citet{Tsai12} calculated the \SFRD\ of sub-regions of NGC~1068 by measuring 8$\mu$m dust continuum flux using box apertures with 3\farcs46 $\times$ 2\farcs56 (242 $\times$ 179 pc$^2$). Assuming Salpeter IMF \citep{Salpeter55}, they measured a total SFR of 3.1 $\pm$ 0.1 \myr\ in the starburst ring of NGC~1068, which is the same as our measurement range. For a fair comparison, we scale the SFR derived in \citet{Tsai12} by Salpeter-to-Kroupa conversion factor 1.6 \citep{Marchesini09} to take into account the difference between the Salpeter IMF and the Kroupa IMF we used. This results in a total SFR of 2.0 $\pm$ 0.1 \myr, which is roughly two-thirds of our total SFR value. Possible reasons for the inconsistency are as follows:

\textbf{Different prescription:} \citet{Tsai12} employed the 8$\mu$m prescription \citep{Wu05} to create their \SFRD\ map. This may lead to underestimating the total SFR because unobscured star-forming regions that can be detected in UV rather than IR are likely missed in the IRAC maps. We also note that the assumed star formation history for 8$\mu$m changes the resultant SFR \citep{KennicuttEvans12}.

\textbf{Different timescale:} Each SFR tracer has a different physical mechanism of emission and therefore traces star-forming regions of different ages and timescales. Our method utilizes emission from ionized gas in \HII\ regions. These emissions represent a nearly instantaneous SFR with the age of $\simeq$3--10~Myr \citep{KennicuttEvans12}. Contrary to ionized gas tracers, the timescale of 8 $\mu$m dust emission exceeds 10~Myr \citep{Calzetti08,KennicuttEvans12}, making it a more suitable tracer for B-stars than for O-stars traced by ionized gas \citep{Peeters04,Calzetti08,Crocker13}. Thus, 8~$\mu$m dust emission is more sensitive to older stars than 
ionized gas. Therefore, the SFR derived from 8~$\mu$m dust emission may be lower than that derived from ionized gas, because the majority of ionizing photons come from massive young stars \citep[e.g.,][]{Jones22}.

Considering the differences in methodology and timescales among studies, we conclude that our SFR measurement represents an instantaneous SFR of the disk of NGC~1068 with more realistic assumptions and corrections.

\subsection{Star-forming activities in NGC~1068}\label{sec:disc:SFactivity1068}

\begin{figure*}[t]
\begin{center}
\includegraphics[width=13cm]{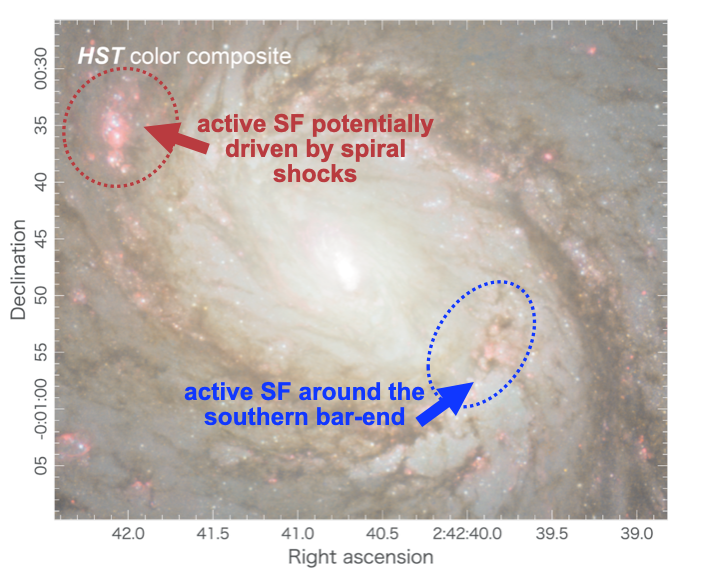} 
\end{center}
\caption{
The {\it HST} color composite map of NGG~1068 with highlights of active star-forming regions. The red ellipse highlights active star-forming regions potentially driven by spiral shocks outer the corotation radius. The blue ellipse highlights active star-forming regions around the southern bar-end.
}\label{Fig:drivenSF_fig15}
\end{figure*}

Figure~\ref{Fig:SFRs_fig13} shows spatial variations of the \SFRD\ within NGC 1068 for cases where the DIG is corrected and uncorrected. The most prominent feature is the high \SFRD\ outside of the bar-end on the starburst ring (the red circle in Figure~\ref{Fig:drivenSF_fig15}). In general, barred spiral galaxies show elevated star formation in bar-ends \citep[e.g.,][]{Maeda23}. Our result can naturally explain the previous NGC~1068 observations (\citealt{Tosaki17,Nakajima23,Sanchez-Garcia22}). Especially, our result supports the findings in \citet{Tosaki17} that significant suppression of a shock tracer CH$_3$OH in the bar-ends of NGC~1068 is due to high-temperature environments of the ISM in the bar-ends caused by active star formation. 

There is a high \SFRD\ blob at the northern tip of the eastern spiral arm structure. We identify \HII\ regions at the same position in the {\it HST} composite map by eye (The red circle in Figure~\ref{Fig:drivenSF_fig15}). Considering the location of the corotation radius (1.4~kpc; \citealt{Schinnerer00}) and the position of spiral shocks \citep{Tosaki17}, we speculate that the molecular gas first flows into the spiral potential well where CH$_3$OH is detected, the gas compression in the potential well leads to subsequent star formation which is visible as the high \SFRD\ blob. A similar mechanism is suggested to explain the star formation at the corotation radius of the nearby spiral galaxy NGC~628 \citep{Herrera20}. Testing this speculation by using our \SFRD\ maps ($=$ \HII\ regions), molecular gas maps ($=$ molecular clouds), stellar mass map ($=$ gravitational potential), and kinematics modeling is a future direction of this work.

\subsection{Comparing electron temperature with other galaxies}\label{sec:disc:comparingTe}

In this section, we compare \Te\ measurements of this study with those of other nearby well-studied galaxies (NGC~253; \citealt{Bendo15} and NGC~4945; \citealt{Bendo16}) utilizing the radio hydrogen recombination line to \ffc\ ratio. We discuss possible factors contributing to differences in \Te\ among these three galaxies.

Metallicity is a possible candidate that causes variations in \Te\ \citep[e.g.,][]{Shaver83,Pilyugin05,Balser24}. Two empirical relations are known to control the metallicity of a galaxy: the mass-metallicity relation (MZR) \citep[e.g.,][]{Kewley01,Curti20} and the negative metallicity gradient \citep[e.g.,][]{Matteucci01,Sanchez-Menguiano16}. MZR, a well-established correlation observed in galaxies (i.e., more massive galaxies tend to have higher metallicities) is likely driven by their star formation histories. The negative gradient within galaxy disks describes the trend observed in many spiral galaxies, where metallicity decreases with increasing distance from the galaxy center, a natural consequence of the inside-out evolution of galaxies \citep[e.g.,][]{Chiappini97,Fu09}.

Table~\ref{table:Te-met} lists the galaxy properties necessary to discuss the origin of the \Te\ variation among the three galaxies. The radius $R_{25}$ is defined as the radius at which the surface brightness of a galaxy in $B$-band reaches 25 mag arcsec$^{-2}$. The ratio $R/R_{25}$ is a dimensionless radius that does not depend on the size of galaxies. It normalizes distances to a galaxy's $R_{25}$, facilitating scale-free analysis of galactic structures and properties across galaxies. In this discussion, $R$ represents the radius of the region within each galaxy where the \Te\ has been measured. $M_{\star}$ expresses the total stellar mass of galaxies. In this discussion, we use \Te\ corrected for the DIG when comparing with NGC~253 and NGC~4945. This is because (1) our DIG-uncorrected \Te\ value for NGC~1068 is unreasonably low, and (2) the radio hydrogen recombination lines used for the SFR measurements for NGC~253 and NGC~4945 are too faint to detect DIG components as like \ffc.

Based on MZR, we expect that the galaxy-averaged metallicity of the three galaxies should be comparable, due to the flat shape of the MZR at more massive regime (See Figure~3 in \citealt{Curti20}). According to \citet{Curti20}, the difference in galaxy-averaged metallicity results in an \Te\ difference of less than 1000~K \citep[See also Figure~25 in][]{Shaver83}. This expected \Te\ difference driven by MZR cannot account for the difference in \Te\ among three galaxies.

Regarding the negative metallicity gradient, $R/R_{25}$ where metallicity is measured in each galaxy is significantly different. High metallicity, typically found near galaxy centers, correlates with lower \Te. NGC~253 has the lowest \Te\ because the measurement is done at the smallest $R/R_{25}$ among the three galaxies. In the case of NGC~1068, the highest \Te\ is measured because $R/R_{25}$ is the largest. Thus, the negative gradient offers a plausible explanation for the differences in \Te\ among the three galaxies. However, it is unclear whether the radial metallicity gradient can fully explain the variations. To verify this, it is necessary to establish a relationship among \Te\ (derived from the ratio of hydrogen recombination line to \ffc), metallicity, and galactocentric radius. To achieve this, we have to verify the conditions under which the \Te\ derived from \ffc\ and \Paa\ shows a good correlation with the \Te\ derived from auroral lines.

\begin{deluxetable*}{cccccc}
\tablecaption{Ionized gas properties of nearby galaxies}\label{table:Te-met}
\tablewidth{0pt}
\tablehead{
\colhead{galaxy name} &
 \colhead{log$_{10}$ $M_{\star}$} & \colhead{\Te}& \colhead{$R/R_{25}$} & \colhead{DIG} & \colhead{tracer combination} \\
\colhead{ } & \colhead{($M_\odot$)} &  \colhead{(K)} & \colhead{ } & \colhead{ } & \colhead{ }\\
\colhead{(1)} & \colhead{(2)} & \colhead{(3)} & \colhead{(4)} & \colhead{(5)} & \colhead{(6)} 
}

\startdata
NGC~253 & 10.64$^{a}$ & $3900\pm300$$^b$ & $\simeq0.002^{\mathrm{b,d}}$ & non-detection & H40$\alpha$ $\&$ free-free. \\
NGC~4945 & 10.36$^{a}$  & $5400\pm600$$^c$ & $\simeq0.02^{\mathrm{c,e}}$ & non-detection & H42$\alpha$ $\&$ free-free. \\
NGC~1068 & 10.91$^{a}$ &  $6690^{+360}_{-340}$& $\simeq0.14$ & corrected & \Paa\ $\&$ free-free. \\
\enddata

\tablecomments{
Column 2: Stellar mass \citep{Leroy19}.
Column 3: Best-fit \Te.
Column 4: Galactocentric radius at which \Te\ is measured and normalized by $R_{25}$.
Column 5: Treatment of the DIG.
Column 6: The combination of the ionized gas tracers used in the SFR calibration method. Note that here we only list measurements employing the combination between the hydrogen recombination line and \ffc.
}
\tablerefs{(a) \citealt{Leroy19}; (b) \citealt{Bendo15}; (c) \citealt{Bendo16}; (d) \citealt{deVaucouleur10}; (e) \citealt{Stanghellini15}.}
\end{deluxetable*}

\section{Summary} \label{sec:summary}

We present SFR measurements based on high-resolution (55~pc) and high-sensitivity observations of the 100~GHz continuum taken by ALMA and the hydrogen recombination line \Paa\ taken by {\it HST} toward the nearby prototypical Seyfert galaxy NGC~1068. 
The main goal of this study is to exploit the synergy between \ffc\ and \Paa\ for deriving reliable SFR for NGC~1068. This specific combination of ionized gas tracers allows us to derive and map accurate \SFRD\ distributions. We obtain the best SFR by calibrating, masking, and correcting the effect of \Te, dust extinction, AGN, and diffuse ionized gas (DIG). The advantage of this synergy is that it compensates for each weakness, i.e., unavoidable dust extinction for the bright \Paa\ and faintness of the extinction-free \ffc. In addition, we demonstrate that consideration for the DIG is crucial for this SFR calibration method. Here we summarize this study:

\begin{itemize}
\item The total SFR obtained by cross-calibrating \ffc\ and \Paa\ is 3.2 $\pm$ 0.5~\myr\ for DIG-corrected case and 9.1 $\pm$ 1.4~\myr\ for DIG-uncorrected case . The most plausible \Te\ is 6690$^{+360}_{-340}$~K for the DIG-corrected case and 3510$^{+120}_{-110}$~K for the DIG-uncorrected case in the starburst ring of NGC~1068. The measurements demonstrate the effects of the DIG on this synergy.
 (Section~\ref{sec:method_result:TeDerive:SFRGuess})

\item 
The DIG-corrected SFR does not match IR-based SFR in the literature. The reason may come from the difference in timescales or the difference in the tracer used in the study (Section~\ref{sec:disc:sfr_tsai}). Our new \SFRD\ maps are consistent with the spatially resolved trends reported in the literature (Section~\ref{sec:disc:SFactivity1068}). 

\item Comparison with studies deriving the total SFR and the galaxy-averaged \Te\ using the same technique (i.e., utilizing a hydrogen recombination line and \ffc) suggests that the difference in \Te\ among the sample galaxies is related to the negative metallicity gradient rather than the mass-metallicity relation (Section~ \ref{sec:disc:comparingTe}). 

\end{itemize}
\clearpage 

\section{acknowledgments}

The authors thank an anonymous referee for comments that improved the contents of this paper. We appreciate M. S{\'a}nchez-Garc{\'\i}a and M. Pereira-Santaella for providing their calibrated {\it HST} \Paa\ map. We acknowledge S. Komugi, T. Michiyama, and K. Muraoka for the useful discussion.
This work was supported by NAOJ ALMA Scientific Research grant No. 2021-18A.
This paper makes use of the following ALMA data: 

ADS/JAO.ALMA\#2011.0.00061.S, 

ADS/JAO.ALMA\#2012.1.00657.S, 

ADS/JAO.ALMA\#2013.1.00055.S, 

ADS/JAO.ALMA\#2013.1.00060.S, 

ADS/JAO.ALMA\#2013.1.00111.S, 

ADS/JAO.ALMA\#2013.1.00188.S, 

ADS/JAO.ALMA\#2013.1.00221.S, 

ADS/JAO.ALMA\#2013.1.00279.S, 

ADS/JAO.ALMA\#2015.1.00960.S, 

ADS/JAO.ALMA\#2016.1.00023.S, 

ADS/JAO.ALMA\#2016.1.00052.S, 

ADS/JAO.ALMA\#2016.1.00232.S, 

ADS/JAO.ALMA\#2017.1.01666.S, 

ADS/JAO.ALMA\#2018.1.00037.S, 

ADS/JAO.ALMA\#2018.1.01506.S, 

ADS/JAO.ALMA\#2018.1.01684.S, 

ADS/JAO.ALMA\#2019.1.00130.S, and 

ADS/JAO.ALMA\#2021.2.00049.S.

ALMA is a partnership of ESO (representing its member states), NSF (USA) and NINS (Japan), together with NRC (Canada), MOST and ASIAA (Taiwan), and KASI (Republic of Korea), in cooperation with the Republic of Chile. The Joint ALMA Observatory is operated by ESO, AUI/NRAO and NAOJ.
This research is based on observations made with the NASA/ESA Hubble Space Telescope, and obtained from the Hubble Legacy Archive, which is a collaboration between the Space Telescope Science Institute (STScI/NASA), the Space Telescope European Coordinating Facility (ST-ECF/ESAC/ESA) and the Canadian Astronomy Data Centre (CADC/NRC/CSA).
This research is based on observations collected at the European Southern Observatory under ESO program 094.B-0321 (A).
The NVAS image was produced as part of the NRAO VLA Archive Survey, (c) AUI/NRAO.
This research has made use of the NASA/IPAC Extragalactic Database (NED), which is funded by the National Aeronautics and Space Administration and operated by the California Institute of Technology. 
Data analysis was in part carried out on the Multi-wavelength Data Analysis System operated by the Astronomy Data Center (ADC), National Astronomical Observatory of Japan. KN acknowledges support from the JSPS KAKENHI Grant Number 19K03937 and the work of KB was partially supported by the JSPS KAKENHI Grant
Number 21K03547 and 23KF0008.
\vspace{5mm}
\facilities{HST (NICMOS), ALMA (Band~3 and Band~6), VLT (MUSE), VLA (Ku-band)}
\software{{\tt ALMA Interferometric Pipeline},
{\tt Astropy} \citep{Astropy13,Astropy18},
{\tt CASA} \citep{CASATeam22},
{\tt NumPy} \citep{Harris20},
{\tt PHANGS-ALMA Pipeline} \citep{Leroy21a},
{\tt SciPy} \citep{Virtanen20},
{\tt spectral-cube} \citep{Ginsburg19},
{\tt radio-beam ({\url{https://pypi.org/project/radio-beam/}}})
}


\clearpage 
\appendix

\section{Extinction correction for \Paa}\label{sec:appendix:Rv}

We correct dust extinction for \Paa\ as described in Section~\ref{sec:method_result:extinctionCorr}. We use the $A_{V}$ map of NGC~1068 produced in \citet{Mingozzi19}. They assumed  the selective extinction \Rvcal\ $=$ 3.1 and \Hb/\Ha\ $=$ 2.86.
We convert \Rvcal\ from 3.1 to 4.05 because of the adaptation to the \Paa\ wavelength. If \Rvcal\ remains at 3.1, the corrected \Paa\ intensity becomes unrealistic for some pixels. We extrapolate \Rvcal\ from 3.1 to 4.05 using the following equation,

\begin{equation}
\EBV\ = \left (\frac{\AV}{\kVcal\ \times 0.44}\right),
\label{eq:}
\end{equation}
where $\EBV$ is the color excess and $\mathrm{\mathrm{\kappa'}} (V)$ is obtained from Equations~2 and 3 of \citet{Calzetti00}.
With the following equations, we get an \Avcal\ map for a different \Rvcal,

%

\begin{eqnarray}
\lefteqn{ E(\beta-\alpha) = } \nonumber \\
&& \left (\frac{\mathrm{\kappa'}(\beta)_{R\mathrm{v}=3.1} - \mathrm{\kappa'}(\alpha)_{R\mathrm{v}=3.1}}{\mathrm{\kappa'}(B)_{R\mathrm{v}=3.1} - \mathrm{\kappa'}(V)_{R\mathrm{v}=3.1}} \right)\times E(B-V)_{R\mathrm{v}=3.1}, \nonumber \\
\lefteqn{ E(B-V)_{R\mathrm{v}=4.05} = } \nonumber \\
&& \left (\frac{\mathrm{\kappa'}(B)_{R\mathrm{v}=4.05} - \mathrm{\kappa'}(V)_{R\mathrm{v}=4.05}}{\mathrm{\kappa'}(\beta)_{R\mathrm{v}=4.05} - \mathrm{\kappa'}(\alpha)_{R\mathrm{v}=4.05}}\right) \times E(\beta-\alpha), \nonumber \\
\lefteqn{ A_{V,R\mathrm{v}=4.05} = } \nonumber \\
&& (\mathrm{\kappa'}(V)_{R\mathrm{v}=4.05} \times 0.44) \times E(B-V)_{R\mathrm{v}=4.05}.
\end{eqnarray}\label{eq:}
We refer the readers to \citet{Calzetti00} for more details. We derive A$_{\lambda 1.875}$/$A_V$ of 0.145 by extrapolating $A_V$ from \Rvcal=3.1 to \Rvcal=4.05. The accuracy of this value is confirmed by its agreement with the value derived by \citet{Piqueras13}, using the observed central flux in the $V$-band.         

\section{Negligible contamination of DIG for \ffc}\label{sec:appendix:contamiDIG}
Here, we demonstrate whether the \ffc\ significantly contains the DIG components in NGC~1068.
First, we calculate the expected emission measure of the DIG ($EM_{\rm DIG}$) based on the \Paa\ luminosity of the DIG components,

\begin{equation}
   \left(\frac{L_{\mathrm{DIG}}}{\mathrm{erg\ s^{-1}}}\right) = \left(\frac{\epsilon}{{\mathrm{erg\ s^{-1}\ cm^{-3}\ /n_en_p}}}\right)\left(\frac{EM_{{\mathrm{DIG}}}}{\mathrm{cm^{-3}}}\right),
   \label{eq:Ldig}
\end{equation}
where $\epsilon$ comes from Equation~\ref{eq:eps}. Then, we measure the expected flux of the \ffc\ based on the \Paa-based $EM_{\rm DIG}$. The equation is,

\begin{equation}
I^{\rm ff}_{\nu} = \frac{2\nu^2}{c^2}k \Te\tau^{\rm ff}_\nu,
\end{equation}
where $I^{\rm ff}_{\nu}$ is the intensity and $\tau^{\rm ff}_\nu$ is the optical depth of \ffc\ which is given by,

\begin{equation}
\tau^{\rm ff}_\nu = 8.235 \times 10^{-2} \Te^{-1.35} \left(\frac{\nu}{\rm GHz}\right)^{-2.1}\left(\frac{EM_{\rm DIG}}{\mathrm{cm^{-6}\ pc}}\right)Z^2,
\end{equation}

where $Z$ is the ionic charge. The typical \Te\ for the DIG in the Milky Way is estimated to be 6000--10000~K, the electron and proton density($n_{\rm e}$, $n_{\rm p}$) to be $\simeq$0.05 $\mathrm{cm^{-3}}$, and the scale height to be $\simeq$1000~$\mathrm{pc}$ (see \citealt{Haffner09} for more details). Based on the calculation described above, the expected peak DIG flux at 100~GHz is 0.01~$\mu$Jy beam$^{-1}$ which is well below the image RMS of our ALMA 100~GHz continuum map (0.1~mJy beam$^{-1}$).

\section{DIG correction to the \Paa}\label{sec:appendix:SIIHa}

\begin{figure*}[t]
 \begin{center}
 \includegraphics[width=15cm]{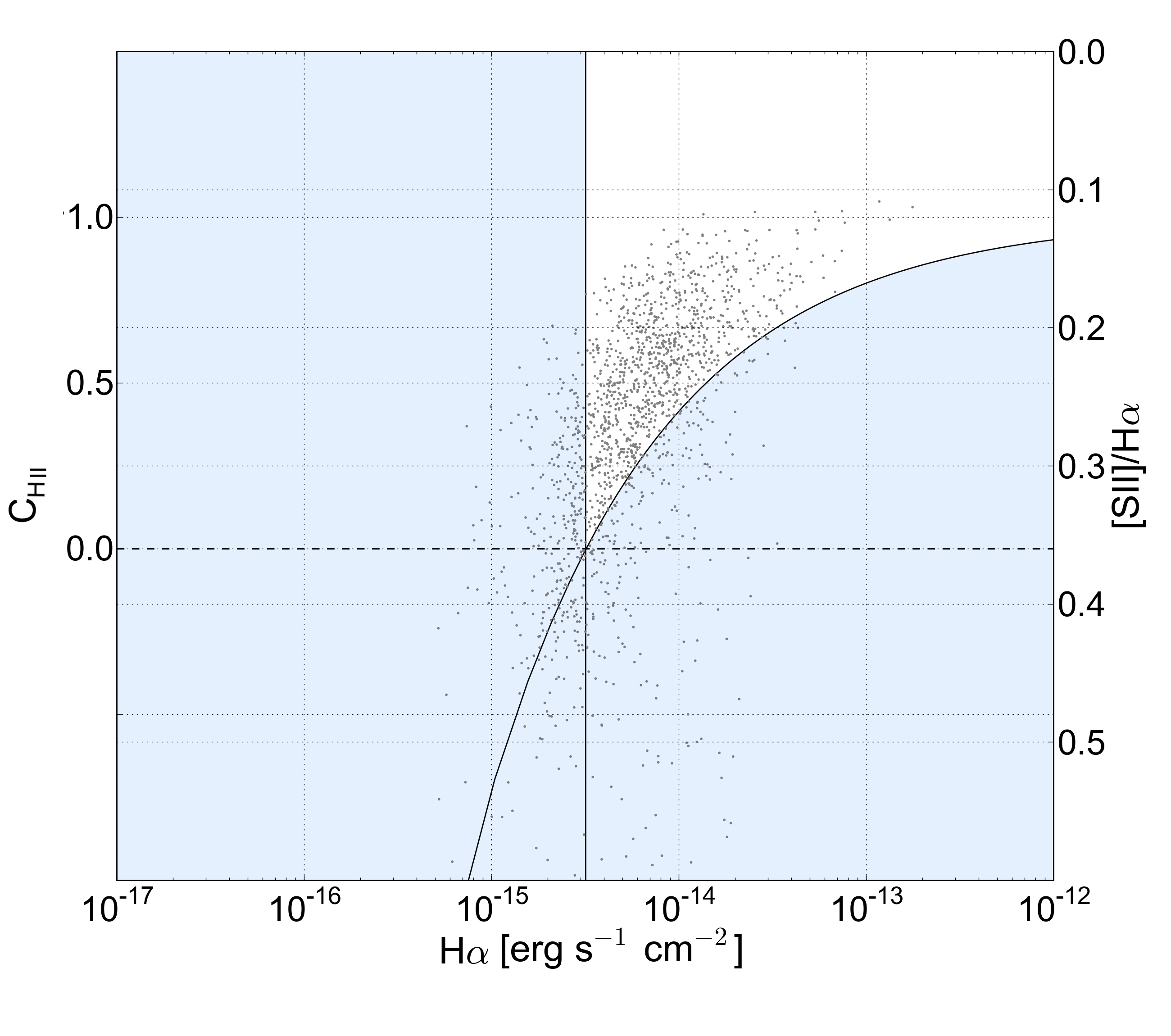} 
\caption{A DIG diagnostic plot with the \SII/\Ha\ method \citep[e.g.,][]{Kaplan16}. The horizontal axis represents the extinction-corrected \Ha\ intensity. The left vertical axis represents the estimated fraction classified as \HII\ region in each data ($C_{\rm HII}$). $C_{\rm{H\,II}}$ below 0.0 indicates DIG contributes 100~\%, and $C_{\rm{H\,II}}$ above 1.0 indicates \HII\ region contributes 100~\% to \Ha\ flux. The right vertical axis represents the \SII/\Ha\ line ratio which is used to calculate $C_{\rm HII}$. The dashed horizontal line indicates the threshold of the absent \HII\ region in the hexagon (i.e., DIG contributes 100 percent of the \Ha\ flux in the hexagon). The black solid curve and vertical line are determined by the observations of Galactic DIG components that typically have low \Ha\ flux and high \SII/\Ha\ line ratio. The minimum value of \Ha\ intensity ($f_0$) at which a \HII\ region can exist in each pixel defined by \citet{Kaplan16} was 3.25 $\times$ 10$^{-15}$ \erg. The data points falling into the blue area of this plot are known to be ``pure DIG". Data points with $>3\sigma$ are displayed.
}\label{fig:siiha_fig21}
\end{center}
\end{figure*}

We use the \SII/\Ha\ method originally proposed by \citet{Blanc09}. This method has been improved to adapt external galaxies by several authors \citep[e.g.,][]{Kaplan16, Kreckel16, Poetrodjojo19}. We use the prescription proposed in \citet{Kaplan16} and \citet{Kreckel16}.
In this method, the data showing the strong intensity of \Ha\ emission are defined as pure \HII\ regions, and \SII/\Ha\ line ratio is used to distinguish low-ionization gas and \HII\ region. Figure~\ref{fig:siiha_fig21} shows a plot necessary to calculate the fraction of DIG at each pixel. The $C_{\rm{H\,II}}$ is the fraction classified as \HII\ region in each data and this is calculated by defining the threshold of \HII\ region and DIG in the \SII/\Ha\ ratio:
\begin{equation}
   \left(\frac{\SII}{\Ha}\right)_{\rm OBS} = C_{\rm{H\,II}} \left(\frac{\SII}{\Ha}\right)_{\rm{H\,II}} + C_{\rm{DIG}} \left(\frac{\SII}{\Ha}\right)_{\rm{DIG}},
   \label{eq:SIIHa1}
\end{equation}
Here, $C_{\rm{H\,II}}$ is the fraction of \Ha\ flux arising from the \HII\ region in each hexagon and $C_{\rm{H\,II}}  = 1 - C_{\rm{DIG}}.$ The minimum of \SII/\Ha\ is used as the threshold for \HII\ regions, and the median of \SII/\Ha\ is the threshold of DIG for a single hexagon on the map \citep{Kreckel16}.

The black curve in Figure~\ref{fig:siiha_fig21} shows the minimum of $C_{\rm{H\,II}}$ for the flux of \Ha. The function of the black curve is defined following:
\begin{equation}
   \left(\frac{\SII}{\Ha}\right) = 1 - \left(\frac{f_0}{f(\Ha)}\right)^{\beta},
   \label{eq:SIIHa2}
\end{equation}
Here, \textit{f}$_0$ is the minimum of the flux of \Ha that can exit \HII\ region, f(\Ha) is the \Ha\ flux, and the exponent $\beta$ is used to accommodate changes in the surface brightness of DIG. 
The minimum value of \Ha\ intensity ($f_0$) at which a \HII\ region can exist in each pixel defined by \citet{Kaplan16} was 3.25 $\times$ 10$^{-15}$ \erg. The data points falling into the blue area of Figure~\ref{fig:siiha_fig21} are known to be ``pure DIG" (see \citealt{Kreckel16} and \citealt{Kaplan16} for more details).

\section{\SFRD\ distribution in NGC~1068}\label{sec:appendix:lognormalhist}

\begin{figure*}[t]
 \begin{center}
  \includegraphics[width=15cm]{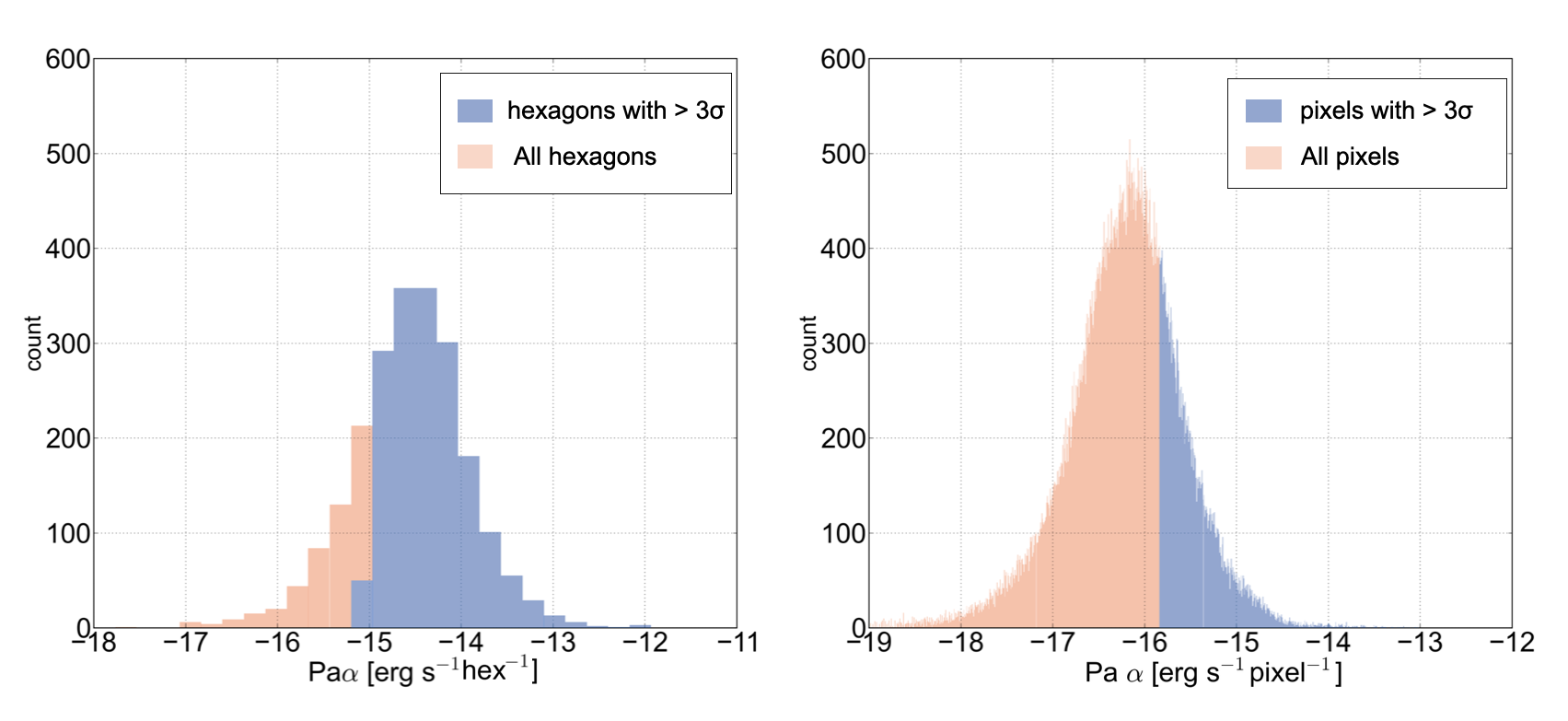} 
 \end{center}
\caption{(left) Histogram of the hexagon-resampled \Paa\ map. The blue bins reflect over 3$\sigma$ level hexagons and the orange bins reflect all hexagons. (right) Histogram of the original \Paa\ map. The blue bins reflect over 3$\sigma$ level pixels and the orange bins reflect all pixels. The data points exceeding 3~$\sigma$ of error are binned.
}\label{fig:lognormal_fig25}
\end{figure*}

The distribution of \SFRD\ in each hexagon is nearly log-normal (Figure~\ref{Fig:SFRh_fig14}). Note that this does not imply a sampling bias. We confirmed this by comparing the histograms of $L_{\rm RL}$ in the pixel bases and hexagon bases as shown in Figure~\ref{fig:lognormal_fig25} and Figure~\ref{Fig:SFRh_fig14} reflect the original distribution of the \SFRD\ ($\propto\, L_{\rm RL}$ ) histogram.

In addition, we explain below why hexagonal sampling increases the proportion of bins exceeding 3$\sigma$ in Figure~\ref{fig:lognormal_fig25}. The RMS value for the hexagon, \( \sigma_{\text{hex}} \), can then be derived through the principle of error propagation as:
\[
\sigma_{\text{hex}} = \sqrt{N} \times \sigma_{\text{pixel}}
\]
Here, the number of data is N and the RMS value for the pixel is \( \sigma_{\text{pixel}} \). Given that a hexagon contains multiple pixels, the signal-to-noise ratio within each hexagon is reduced. Consequently, post-sampling into hexagons, an increase appears in the fraction of bins that exhibit statistically significant signals (exceeding 3 $\sigma$). Therefore, by employing hexagonal sampling, we leverage the benefit of effectively enhancing the signal-to-noise ratio.

\section{Spatial distribution of electron temperature}\label{sec:appendix:spatialTe}

The reason for assuming a single \Te\ to derive SFR in Section~\ref{sec:method_result:TeDerive:TeGuess} is to take advantage of the number of data points for the fitting. Here we examine whether the other manner works, i.e., deriving \Te\ at each hexagon by using a smaller number of data points. We use the same fitting method described in Section~\ref{sec:method_result:TeDerive:TeGuess}, but we perform the fitting for each hexagon with at least 6 surrounding hexagons. This method allows us to create an \Te\ map. This effectively degrades the spatial resolution by a factor of two compared to the original maps with 55~pc resolution. The detailed procedure for the fitting is as follows: (1) we select hexagons with at least 6 adjacent hexagons detected in both \Paa\ and \ffc, (2) the fitting is performed for the 6 or 7 data points to derive a single \Te, and (3) the obtained \Te\ is assigned to the central hexagon (i.e., the neighboring hexagons assist the fitting).

\begin{figure*}[th]
\centering  \includegraphics[width=0.7\textwidth]{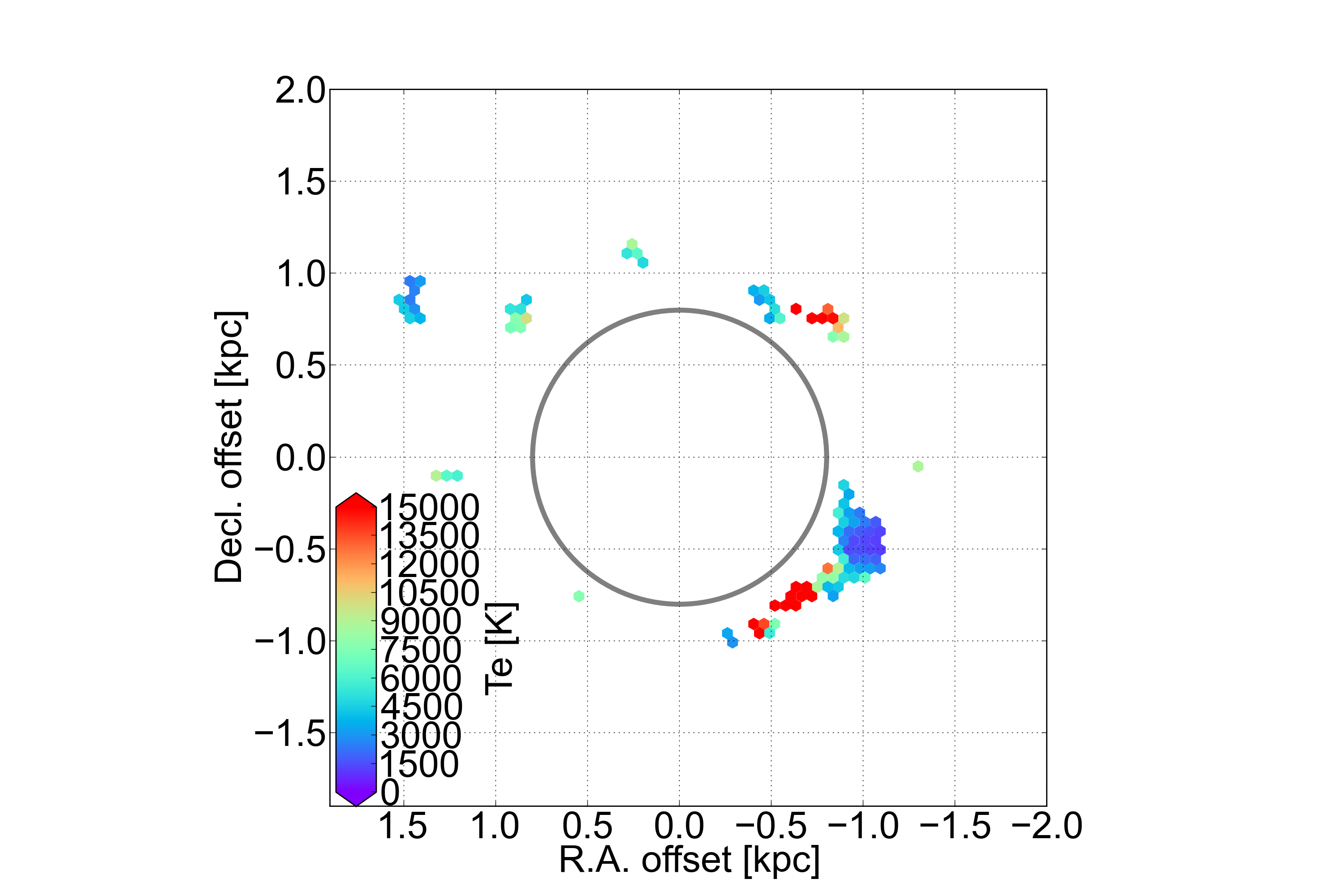} 
\caption{\Te\ map derived by fitting a small number of data points for each hexagon. Pixels with robust \Te\ measurements are only displayed. The black circle highlights the central r$\sim$0.8~kpc region which is severely affected by the AGN jet and outflow \citep[e.g.,][]{Garcia-Burrillo14,Saito22b}. Data points with $>3\sigma$ are displayed.}
\label{fig:Tem_fig16}
\end{figure*}

\begin{figure*}[th]
 \centering \includegraphics[width=0.5\textwidth]{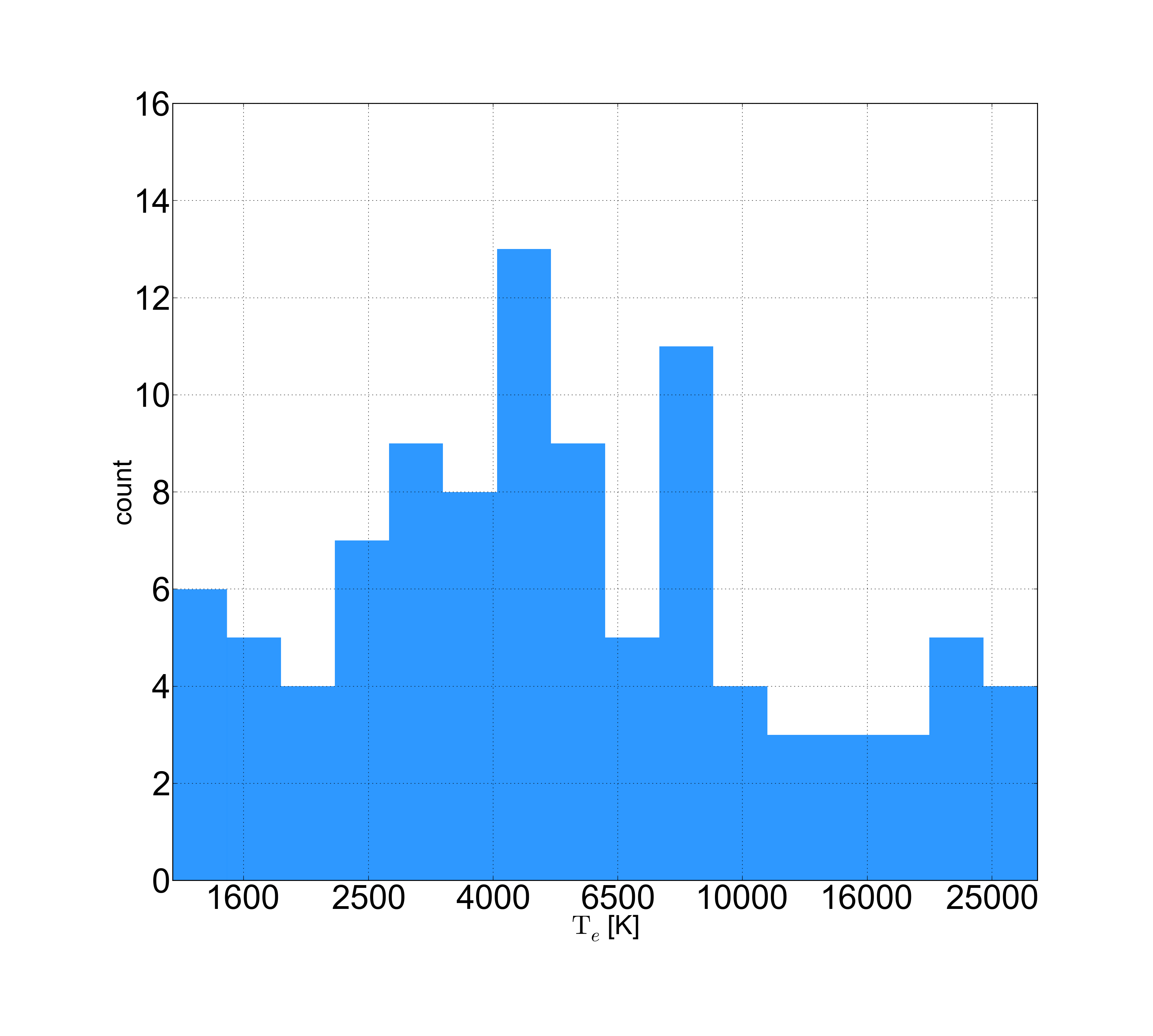}
  \caption{Histogram of the \Te\ map shown in Figure~\ref{fig:Tem_fig16}. The 16$^{\rm th}$--50$^{\rm th}$--84$^{\rm th}$ percentiles are 2410--4630--12850~K.}
  \label{fig:Teh_fig17}
\end{figure*}

The map and histogram of \Te\ based on this method are shown in Figure~\ref{fig:Tem_fig16} and Figure~\ref{fig:Teh_fig17}, respectively. In the \Te\ map, hexagons greater than three times the fitting error are displayed. \Te\ of the typical \HII\ region is around 5,000 and 14,000~K \citep[e.g.,][]{Draine11}. However, as shown in Figures~\ref{fig:Tem_fig16} and \ref{fig:Teh_fig17}, the \Te\ distribution based on our method shows a wider range above 15,000 K or below 4,000 K. These extreme values are theoretically difficult to explain. This may be due to the smaller number of available data points for the fitting at each position. This result affirms the advantage of estimating a single \Te\ in the starburst ring.

\bibliography{sample631}{}
\bibliographystyle{aasjournal}

\end{document}